\documentclass[11pt]{article}
\DeclareMathAlphabet{\scr}{U}{rsfs}{m}{n}
\usepackage{latexsym}
\usepackage[mathscr]{eucal}
\usepackage{amsfonts}
\usepackage{amscd}
\usepackage{cite}
\usepackage{amssymb}
\usepackage[centertags]{amsmath}
\usepackage{dsfont}
\usepackage{yfonts}
\usepackage{enumerate}
\usepackage{epsfig}
\usepackage{amsmath}
\usepackage{graphicx}
\usepackage{bbold}
\usepackage{bbm}
\usepackage[english]{babel}

\newcommand{\cleqn}{\setcounter{equation}{0}}
\def\e{\epsilon}

\def\La{\Lambda}

\def\n{\nu}

\newcommand{\newc}{\newcommand}
\newc{\be}{\begin{equation}}
\newc{\ee}{\end{equation}}
\newc{\ben}{\begin{equation*}}
\newc{\een}{\end{equation*}}
\newc{\bea}{\begin{eqnarray}}
\newc{\eea}{\end{eqnarray}}
\newc{\bean}{\begin{eqnarray*}}
\newc{\eean}{\end{eqnarray*}}
\newc{\ol}{\overline}
\newc{\wt}{\widetilde}
\newc{\bs}{\boldsymbol}
\newc{\m}{\mathcal}
\newc{\lra}{~\longrightarrow~}

\newc{\VEV}[1]{\langle #1 \rangle}

\newcommand{\betabeta}{\mbox{$(\beta \beta)_{0 \nu}  $}}

\newcommand{\dmsol}{\mbox{$\Delta m^2_{\odot}$}}
\newcommand{\dma}{\mbox{$\Delta m^2_{\rm A}$}}

\newcommand{\gsim}{\lower.7ex\hbox{$\;\stackrel{\textstyle>}{\sim}\;$}}
\newcommand{\lsim}{\lower.7ex\hbox{$\;\stackrel{\textstyle<}{\sim}\;$}}

\newcommand{\GeV}{\ensuremath{\:\mathrm{GeV}}}

\begin{document}

\begin{titlepage}
\begin{flushright}
SISSA 69/2009/EP
\end{flushright}
\vspace*{5mm}

\begin{center}
{\Large\sffamily\bfseries
\mathversion{bold}
Charged Lepton Flavour Violating
Radiative Decays  $\ell_i \to \ell_j + \gamma$ \\
in See-Saw Models with $A_4$ Symmetry
\mathversion{normal}} \\[13mm]
{\large
C.~Hagedorn~$^{a,}$\footnote{E-mail: \texttt{hagedorn@sissa.it}},
E.~Molinaro~$^{a,}$\footnote{E-mail: \texttt{molinaro@sissa.it}}
and S.~T.~Petcov~$^{a,b,}$\footnote{Also at: Institute of Nuclear
Research and Nuclear Energy,
Bulgarian Academy  of Sciences\\1784 Sofia, Bulgaria.}
}
\\[5mm]
{\small \textit{
$^a$ SISSA and INFN-Sezione di Trieste\\Trieste I-34151, Italy\\
$^b$IPMU, University of Tokyo, Tokyo, Japan}}
\vspace*{1.0cm}
\end{center}
\normalsize
\begin{abstract}
The charged lepton flavour violating (LFV) radiative decays,
$\mu\rightarrow e+\gamma$,
$\tau\rightarrow \mu+\gamma$ and
$\tau\rightarrow e +\gamma$
are investigated in a class of supersymmetric
$A_4$ models with three heavy right-handed (RH) Majorana neutrinos,
in which the lepton (neutrino) mixing is
predicted to leading order (LO) to be tri-bimaximal.
The light neutrino masses are generated
via the type I see-saw mechanism.
The analysis is done within
the framework of the minimal
supergravity (mSUGRA) scenario,
which provides flavour universal boundary conditions
at the scale of grand unification
$M_X \approx 2 \times 10^{16}$ GeV.
Detailed predictions for the rates of the three
LFV decays are obtained in two explicit realisations
of the $A_4$ models
due to Altarelli and Feruglio and Altarelli and Meloni, respectively.
\end{abstract}
\end{titlepage}
\setcounter{footnote}{0}

\mathversion{bold}
\section{Introduction}
\mathversion{normal}

 Neutrino experiments revealed a
totally different mixing pattern in the lepton sector
compared to the one observed in the quark sector
and encoded in the CKM matrix.
Indeed, two lepton (neutrino) mixing angles
are large \cite{nudata},
\begin{equation}
\sin ^2 \theta_{12} = 0.304 ^{+0.066} _{-0.054} \;\;\; \mbox{and} \;\;\;
\sin ^2 \theta_{23} = 0.50 ^{+0.17} _{-0.14} \;\;\; (3 \sigma) \; ,
\end{equation}
%
while the third one, the CHOOZ
angle, is limited by the upper bound \cite{nudata}:
\begin{equation}
\sin^2 \theta_{13} < 0.056\, ~~~ (3 \sigma) ~~.
\end{equation}
%
The mixing
observed in the lepton sector is
remarkably close to the tri-bimaximal (TB) one
proposed in \cite{hps}:
\begin{equation}
\label{UTBM}
U_{TB} = \left(
\begin{array}{ccc}
        \sqrt{2/3}  & 1/\sqrt{3} & 0\\
        -1/\sqrt{6} & 1/\sqrt{3} & 1/\sqrt{2}\\
        -1/\sqrt{6} & 1/\sqrt{3} & -1/\sqrt{2}
\end{array}
\right)
 \; .
\end{equation}
%
In the case of TB mixing,
the Pontecorvo-Maki-Nakagawa-Sakata (PMNS)
neutrino (lepton) mixing matrix $U$
is given, in general, by:
\begin{equation}
\label{UPMNS}
U = U_{TB}  \left(
\begin{array}{ccc}
        1  & 0 & 0\\
        0 & e^{i\frac{\alpha_{21}}{2}} & 0\\
        0 & 0 & e^{i\frac{\alpha_{31}}{2}}
\end{array}
\right)
 \; ,
\end{equation}
%
where $\alpha_{21}$ and $\alpha_{31}$
are two Majorana CP violating phases \cite{BHP80}.
The three neutrino mixing angles in the
standard parametrisation of the
PMNS matrix (see, e.g. \cite{UPMNSstandard})
in this case have fixed values
determined by:
\begin{equation}
\sin ^2 \theta_{12}^{TB} = 1/3 \; , \;\;
\sin ^2 \theta_{23}^{TB} = 1/2
\;\;\; \mbox{and} \;\;\;
\sin ^2 \theta_{13}^{TB} = 0 \; .
\end{equation}
%
Thus, in the case of exact TB neutrino mixing,
the Dirac CP violating phase is not present in the
PMNS matrix and the CP symmetry can be violated only
by the Majorana phases.

  It has been shown in \cite{A4models,AF1,AF2,AM}
that TB mixing can arise in a class of models
in which  the alternating (tetrahedral) group
$A_4$ serves as flavour symmetry group
\footnote{TB mixing can also be derived from
models with an $S_4$ flavour symmetry \cite{S4}.}.
The light neutrino masses are generated in
these models either through dimension-5 operators or through
the type I see-saw mechanism. In the following we concentrate on
models which contain three right-handed (RH) neutrinos and employ
the latter mechanism for neutrino mass generation.
The light massive neutrinos are Majorana particles.
The $A_4$ symmetry is spontaneously broken at some
high energy scale by the vacuum expectation
values (VEVs) of a set of scalar fields,
called flavons, which transform trivially
under the gauge symmetry group of the model.
In this approach, the TB mixing appears at leading
order (LO). The LO results get corrected
by higher order terms which can
play important role
in the phenomenology of the models.
Such corrections are suppressed, in general,
by a factor $\epsilon$ with respect to
the LO results, where
$\epsilon$ denotes the ratio of a generic
flavon VEV and the cutoff scale $\La$ of the theory.
The scale $\La$ is assumed to be close to the
scale of grand unification. The parameter
$\epsilon$ typically takes values
between a $\mbox{few} \times 10^{-3}$ and $0.05$.

  Apart from predictions for the mixing angles,
in $A_4$ models one can obtain also predictions
for the light and heavy Majorana neutrino
mass spectrum \cite{AF2,AM,HMP09} and
constraints on the Majorana CP violating
phases. More specifically, both types of light
neutrino mass spectrum - with normal and
inverted ordering, are possible, but
in both cases the lightest neutrino mass
is constrained to lie in certain
intervals. As a consequence, one has specific
testable predictions for the
magnitudes of the sum of neutrino
masses and of the effective Majorana mass,
measured in neutrinoless double beta ($\betabeta$-) decay
(see, e.g. \cite{UPMNSstandard,STPFocusNu04}).
Recently, it has been shown that the observed baryon
asymmetry of the Universe can be
generated in the $A_4$ models under discussion
through the leptogenesis mechanism
\cite{HMP09} (see also \cite{BDiBFN09};
for earlier discussions see  \cite{leptoA4early}).
The CP (lepton charge) asymmetry, leading to
the requisite baryon asymmetry,
is proportional to
the forth power of the $A_4$ flavour symmetry
breaking parameter $\epsilon$ \cite{HMP09}.
The observed relatively small
value of the baryon asymmetry
\cite{YBexp}
is thus to large extent due
to the smallness of
\footnote{The necessary additional suppression
is provided by washout effects \cite{HMP09}.}
$\epsilon$.
Furthermore, the sign of one of the fundamental
parameters of the model, which determines
the light and the heavy neutrino mass spectrum,
can be uniquely fixed by the requirement that the
generated baryon asymmetry has the correct sign \cite{HMP09}.

  Other phenomenological predictions of
this type of $A_4$ models for, e.g. the rates of charged
lepton flavour violating (LFV) radiative decays
$\ell_i \to \ell_j + \gamma$, $i > j$,
$i,j=1,2,3$,
where $\ell_1\equiv e$, $\ell_2\equiv \mu$,
$\ell_3\equiv \tau$,
the electric dipole moments (EDMs)
and magnetic dipole moments (MDMs)
of the charged leptons,
have been studied
using effective field theory methods in \cite{EFTA4}.
In this approach \cite{EFTA4},
a new physics scale $M$ is assumed
to exist at $(1 \div 10)$ TeV.
The effective dimension-6 operator
mediating the charged
LFV radiative decays and
generating new contributions to charged
lepton EDMs and MDMs, is suppressed by this
scale $M$, instead of the high energy scale $\La$.
Thus, the rates of the LFV decays $\ell_i \to \ell_j + \gamma$
and the EDM of the electron, $d_e$,
can have values close to and even above the existing experimental
upper limits
\footnote{As is well known,
in the minimal extension of the Standard Model (SM) with massive
neutrinos and neutrino mixing, the rates and cross sections of the LFV
processes are suppressed by the factors~\cite{SP76} (see
also~\cite{BPP77}) $2.2\times 10^{-4} \, |U_{ej}m^2_{j}U^{*}_{\mu j}|^2/M_W^4
\lesssim 5.2 \times 10^{-48}$, $M_W$, $m_j$ and $U_{lj}$, $l=e,\mu$, being
the $W^{\pm}$ mass, light neutrino masses and
elements of the PMNS matrix. This
renders the LFV processes unobservable.}.
Assuming that the
flavour structure of the indicated dimension-6
operator is also determined by the $A_4$ symmetry,
one finds that its form in flavour space is similar
to the one of the charged
lepton mass matrix. In \cite{EFTA4}
the dependence of the branching ratios
$B (\ell_i \to \ell_j + \gamma)$,
the EDMs and MDMs on $\epsilon$
has been analyzed in detail. It was
found that the contributions of
the new physics to the EDMs and MDMs
arise at LO in $\epsilon$,
whereas the LFV transitions
are generated only at next-to-leading order (NLO).
It was shown that
$B (\ell_i \to \ell_j + \gamma)$ scales
as $\epsilon^2$, independently of the
type of the decaying lepton. Correspondingly,
all charged LFV radiative decays
are predicted to have similar branching ratios.
The existing stringent experimental upper
bound on $B (\mu \to e + \gamma)$
can be satisfied if
the new physics scale $M > 10$ TeV
\footnote{It was indicated in \cite{EFTA4}
that the experimental upper limit on
the EDM of the electron leads to the
constraint $M> 80$ TeV, if the
CP violating phases associated with the new
physics at the scale $M$
have generic values of order one.}.
These results were shown to be independent of the
generation mechanism of the light neutrino masses.

   In this paper we follow a different approach
and compute the branching ratios of
charged LFV radiative decays in two $A_4$ models
which are based on the Minimal Supersymmetric extension
of the Standard Model (MSSM) with
three RH Majorana neutrinos
(for an extensive list of articles in which
the charged LFV radiative decays
were studied see \cite{LFVreview,BiPet87}).
We choose as framework the minimal
supergravity (mSUGRA) scenario,
which provides flavour universal boundary conditions
at the scale of grand unification
$M_X \approx 2 \times 10^{16}$ GeV
\footnote{As has recently been
discussed, in the context of global supersymmetry (SUSY) \cite{VEVsaux}, the presence of
soft SUSY breaking terms in the flavon
sector can lead to additional flavour
non-universal contributions to
the sfermion soft masses. These vanish in
the limit of universal soft SUSY terms
in the flavon potential.}.
The SUSY breaking and
the sparticle masses are
completely specified
by the flavour universal
mass parameters $m_0$, $m_{1/2}$ and $A_0$.
As is well known, this class of models contains
also two low energy quantities $\tan\beta$
and $\mbox{sign}(\mu)$ as free parameters.
Off-diagonal elements in the slepton mass matrices,
which can lead to relatively large branching
ratios of the LFV decays
$\ell_i \to \ell_j + \gamma$,
are generated through renormalization group (RG) effects
associated with the three heavy RH Majorana neutrinos
\cite{BorzMas86}. In the present article
the branching ratios $B (\ell_i \to \ell_j + \gamma)$
are calculated using the analytic approximations
developed in \cite{Hisano96,BR1,BR2}.
In this approach $B (\ell_i \to \ell_j + \gamma)$
depend only on the generated off-diagonal elements of
the mass matrix of the left-handed sleptons,
which are functions of the
matrix of the neutrino Yukawa couplings
(or of the neutrino Dirac mass matrix),
of the three RH Majorana
neutrino masses, and
of the  SUSY breaking parameters
$m_0$, $m_{1/2}$ and $A_0$ (see below).
We discuss
the case of rather generic NLO
corrections to the neutrino Dirac mass matrix.
The numerical results
we present are obtained for two
explicit realizations of $A_4$ models - those
by Altarelli and Feruglio (AF) \cite{AF2} and
by Altarelli and Meloni (AM) \cite{AM}.
In contrast to the results found in the
effective field theory approach in \cite{EFTA4},
where the setup of the AF model was considered,
the branching ratios $B (\ell_i \to \ell_j + \gamma)$
depend in the case we analyse on the
type of the light neutrino mass spectrum,
and do not necessarily scale with the symmetry
breaking parameter $\epsilon$.

From the analytical study of the decay
rates we can infer, for instance,
that in the case of light neutrino mass spectrum with
normal ordering (NO), all branching ratios
are predicted to be independent of the
symmetry breaking parameter $\epsilon$, in
contrast to the results found in the effective
field theory approach in \cite{EFTA4}.
The suppression of the branching ratios
$B (\ell_i \to \ell_j + \gamma)$
in our approach is thus a consequence of
the off-diagonal elements of
the slepton mass matrices being generated
through RG effects and not
of the smallness
of $\epsilon$.
A crucial difference
between  the predictions for
$B(\ell_i \to \ell_j + \gamma)$
of the AF and the AM models
is related to the fact that in
the AF model there exists an
additional source of suppression
of $B(\ell_i \to \ell_j + \gamma)$.
This suppression
is due to the specific pattern of the
NLO corrections (or equivalently due to
the specific form of the VEVs of the flavons
contributing at NLO to the neutrino Dirac
mass matrix) in the AF model.
As a consequence, the predictions for
the branching ratios of $\mu\to e +\gamma$ and $\tau\to e +\gamma$
decays  in the case of light neutrino mass spectrum
with inverted ordering (IO) in the AF model
differ significantly from those
obtained in the AM model.

  In the numerical study we perform we obtain predictions
for the $\ell_i \to \ell_j + \gamma$
decay branching ratios in the AF and AM models
for one specific point in the mSUGRA parameter space,
which is compatible with direct bounds on
sparticle masses and the requirement of
having a viable dark matter (DM) candidate
and the necessary amount of DM.
Results for other points in the   mSUGRA parameter space can be easily
derived by modifying a certain rescaling function
and adjusting $\tan\beta$, present in the expressions for
$B(\ell_i \to \ell_j + \gamma)$.
We discuss also the possibility of having
$B(\mu \to e + \gamma) > 10^{-13}$
and $B(\tau \to \mu + \gamma) \approx 10^{-9}$
in the case of a relatively light sparticle mass spectrum,
which is easily accessible to LHC
\footnote{For recent calculations of the
rates of charged LFV decays
and $\mu - e$ conversion rates
within the mSUGRA framework with
three RH neutrinos and $SO(10)$-inspired
fermion mass matrices,
in which also the DM constraints are satisfied,
see e.g. \cite{Barger}.}. We also show plots for
specific values of the parameters of an $A_4$ model with
generic NLO corrections, which illustrate aspects of
the predictions for $B(\mu \to e + \gamma)$
that cannot be seen in the scatter plots.

 The paper is organized as follows:
in Section 2 we present the aspects of
the two $A_4$ models, relevant for our study
and discuss the generic structure
of the NLO corrections to the neutrino Dirac
mass matrix. In Section 3 we
identify the LO and NLO
contributions in $B(\ell_i \to \ell_j + \gamma)$
in the AF and AM models.
This is done within the framework of mSUGRA
and for the two possible
types - with normal or inverted ordering,
of the light neutrino mass spectrum.
In Section 4 we present results of the
numerical calculations of
the branching ratios $B(\ell_i \to \ell_j + \gamma)$.
We briefly comment
on results for $\mu - e$ conversion and
the decays $\ell_i \to 3 \ell_j$
in Section 5. Finally,
Section 6 contains a summary of the results
obtained in the present work and conclusions.

\mathversion{bold}
\section{$A_4$ Models}
\mathversion{normal}

The $A_4$ models \cite{AF2,AM} we discuss in this paper share several
features: the left-handed lepton doublets $l$
and the three RH neutrinos $\nu^c$ transform
as triplets under $A_4$. In contrast, the right-handed
charged lepton fields
$e^c$, $\mu^c$ and $\tau^c$ are singlets under $A_4$
\footnote{In the AF model they transform as the
three inequivalent one-dimensional representations $1$, $1'$ and $1''$,
whereas in the AM model all three right-handed charged lepton fields
transform trivially under $A_4$.
This, however, is irrelevant for our discussion here.}.
The Majorana mass matrix $m_M$ of the RH neutrinos
is generated through the couplings:
\begin{equation}
a \xi (\nu^c \nu^c) + b \, (\nu^c \nu^c \varphi_S)
\end{equation}
%
where $(\cdots)$ denotes the contraction to an $A_4$ invariant
\footnote{There might exist an additional direct mass term,
as in the AM model, $M (\nu^c \nu^c)$. However, this term leads to
the same contribution as the term $\xi (\nu^c\nu^c)$.}
and $\varphi_S \sim 3$ and $\xi \sim 1$ under $A_4$.
Here and in the following we adopt the convention
for the group $A_4$  as given in
\cite{AF2,AM}. The vacuum alignment of
$\xi$ and $\varphi_S$ achieved, e.g. in \cite{AF2,AM},
is given by:
\begin{equation}
\langle \varphi_S \rangle = v_S \, \epsilon \, \La (1,1,1)^t
\;\;\; \mbox{and} \;\;\;
\langle \xi \rangle = u \, \epsilon \, \La
\end{equation}
%
where $v_S$ and $u$ are assumed to be complex numbers having
an absolute value of order one.
The (real and positive) parameter $\epsilon$
is associated with the ratio of a typical VEV
of a flavon and the cutoff scale $\La$ of the theory.
The generic size of $\epsilon$ is around $0.01$.
At the end of this section we will specify the range
of $\epsilon$ in greater detail.
The matrix $m_M$ can be parametrized as:
\begin{equation}
m_M = \left( \begin{array}{ccc}
        X + 2 Z & -Z & -Z\\
        -Z & 2 Z & X - Z\\
        -Z & X - Z & 2 Z
\end{array}
\right) \; .
\end{equation}
%
It contains two complex parameters $X$ and $Z$
which are conveniently expressed
through their ratio $\alpha=|3Z/X|$,
their relative phase $\phi=\arg(Z)-\arg(X)$ and $|X|$.
The parameter $|X|$ determines the
absolute mass scale of the RH neutrinos.
The matrix $m_M$ is diagonalized
by $U_{TB}$ so that:
\begin{equation}
\hat{U}_{TB} = U_{TB} \, \Omega \;\;\; \mbox{with} \;\;\; \Omega = \mbox{diag} (e^{-i \varphi_1/2}, e^{-i \varphi_2/2}, e^{-i \varphi_3/2})\label{UTB}
\end{equation}
%
leads to
\begin{equation}
\hat{U}_{TB}^T m_M \hat{U}_{TB} = \mbox{diag} (M_1, M_2, M_3)\,,
\end{equation}
%
$M_i$ being the physical RH neutrino masses.
It has been shown in \cite{HMP09} that one can set $\varphi_1=0$ without loss
of generality. Then the phases $\varphi_2$ and $\varphi_3$ coincide at LO in the expansion parameter $\epsilon$
with the low energy Majorana phases $\alpha_{21}$ and $\alpha_{31}$
as defined in eq. (\ref{UPMNS}).

\indent The neutrino Yukawa couplings
in the models considered read:
\begin{equation}
\label{LOmd}
y_\nu (\nu^c l) h_u
\end{equation}
%
so that the neutrino Dirac mass matrix has the simple form:
\footnote{We use the convention in which the RH neutrino
fields are on the left-hand side of the mass matrix
and the left-handed fields are on the right-hand side.}
\begin{equation}
m_D = y_\nu \left( \begin{array}{ccc}
        1 & 0 & 0\\
        0 & 0 & 1\\
        0 & 1 & 0
\end{array}
\right) \, v_u\label{mD}
\end{equation}
%
where $v_u$ denotes the VEV of the MSSM Higgs doublet $h_u$.
We can define the matrix of neutrino Yukawa couplings as:
\begin{equation}
Y_\nu = \frac{m_D}{v_u} \; .\label{YnuOr}
\end{equation}
%
The light neutrino mass matrix
arises from the type I see-saw mechanism:
\begin{equation}
m_\nu=-m_D^T m_M^{-1} m_D \; .
\end{equation}
%
It is diagonalized by $U_{TB}$. The light neutrino masses
$m_i$, $i=1,2,3$, are given by:
\begin{equation}
m_i = \frac{y_\nu^2 v_u^2}{M_i} \;.
\label{numass}
\end{equation}
%
At LO, the charged lepton mass matrix $m_l$ is diagonal in these
models. In the AF model
the charged lepton masses are generated by
the coupling to the flavon $\varphi_T$ with its
alignment $\langle\varphi_T\rangle \propto (1,0,0)^t$
(and the coupling to a Froggatt-Nielsen field),
whereas in the AM model they appear due to the couplings
with the flavons $\varphi_T$ and $\xi'$ having the alignments
$\langle\varphi_T\rangle \propto (0,1,0)^t$ and
$\langle \xi'\rangle \neq 0$.
Note, in particular, that the mass of the
$\tau$ lepton stems from a
non-renormalizable coupling:
\begin{equation}
y_\tau (\tau^c l \varphi_T) h_d/\La \; .
\end{equation}
%
Since $m_l$ is diagonal at this level,
the lepton mixing originates only from
the neutrino sector and is
given by eq. (\ref{UPMNS}).

  This LO result gets corrected by multi-flavon insertions,
as well as by shifts in the VEVs of the flavons.
As a consequence, the matrices $m_M$, $m_D$ and $m_l$
receive corrections. Correspondingly, the lepton masses and
mixings receive relative corrections of order $\epsilon$.
For our study of LFV decays, the form of
the corrections of the neutrino
Yukawa couplings is of special interest.
Instead of discussing these for the two specific
models, the AF and the AM models,
we give here a general parametrization of
the form of these corrections. We start by
writing down the co-variants for the case $\nu^c \sim 3$ and $l \sim 3$
under $A_4$:
\begin{eqnarray}
(\nu^c l) = \nu^c_1 l_1 + \nu^c_3 l_2 + \nu^c_2 l_3 &\sim& 1\\
(\nu^c l)' =\nu^c_3 l_3 + \nu^c_2 l_1 + \nu^c_1 l_2 &\sim& 1'\\
(\nu^c l)'' =\nu^c_2 l_2 + \nu^c_3 l_1 + \nu^c_1 l_3 &\sim& 1''\\
(\nu^c l)_S =\left( \begin{array}{c}
        2 \nu^c_1 l_1 - \nu^c_3 l_2 - \nu^c_2 l_3\\
        2 \nu^c_3 l_3 - \nu^c_2 l_1 - \nu^c_1 l_2\\
        2 \nu^c_2 l_2 - \nu^c_3 l_1 - \nu^c_1 l_3
\end{array}
\right) &\sim& 3_S\\
(\nu^c l)_A =\left( \begin{array}{c}
        \nu^c_3 l_2 - \nu^c_2 l_3\\
        \nu^c_2 l_1 - \nu^c_1 l_2\\
        - \nu^c_3 l_1 + \nu^c_1 l_3
\end{array}
\right) &\sim& 3_A
\end{eqnarray}
%
where $3_{S (A)}$ is the (anti-)symmetric triplet
in the product $3 \times 3$. As one can see, the
structure of $m_D$ at LO coincides with the
structure coming from the $A_4$ invariant.

  We shall discuss first the contributions
which arise at the NLO level through
multi-flavon insertions.
We assume that such contributions
arise at the level of one flavon
insertions and are thus suppressed by $\epsilon$ relative
to the LO result. This is true in the
two realizations which we study below numerically
\footnote{One could also imagine models in which
such contributions are suppressed stronger by higher powers of
$\epsilon$. However, in the majority of the models, the
NLO corrections are suppressed by $\epsilon$ only
compared to the LO result.}.
All NLO contributions which
are of the same form as the
LO result can be simply absorbed into the latter.
Contributions which cannot be absorbed give
rise to NLO terms of the form:
\begin{equation}
y_{1'}^\nu (\nu^c l)' \psi'' h_u/\La    + y_{1''}^\nu (\nu^c l)'' \psi' h_u/\La
+ y_{S}^\nu (\nu^c l)_S \phi h_u/\La +  y_{A}^\nu (\nu^c l)_A \phi h_u/\La
\end{equation}
%
where $\psi'$ and $\psi''$ stand for
flavons which transform as $1'$ and $1''$ under $A_4$,
respectively. Here $\phi$ denotes a triplet
under $A_4$ and, for simplicity,
we assume that there is only one such contribution.
For $\langle \psi'\rangle= w'\, \epsilon \,\La $,
$\langle \psi''\rangle= w'' \, \epsilon \,\La$
and $\langle\phi\rangle = (x_1,x_2,x_3)^t \, \epsilon \,\La$
(with $w'$, $w''$ and $x_i$ being complex numbers whose
absolute value is of order one)
we find that these induce matrix structures of the type:
\begin{eqnarray}\label{deltamD}
\delta m_D &=&
y_{1'}^\nu w'' \, \epsilon \left( \begin{array}{ccc}
        0 & 1 & 0\\
        1 & 0 & 0\\
        0 & 0 & 1
\end{array}
\right) \, v_u +
y_{1''}^\nu w' \, \epsilon \left( \begin{array}{ccc}
        0 & 0 & 1\\
        0 & 1 & 0\\
        1 & 0 & 0
\end{array}
\right)  \, v_u  \\ \nonumber
&& + y_{S}^\nu \, \epsilon \left( \begin{array}{ccc}
        2 x_1 & -x_3 & -x_2\\
        -x_3 & 2 x_2 & -x_1\\
        -x_2 & -x_1 & 2 x_3
\end{array}
\right)  \, v_u +
y_{A}^\nu \, \epsilon \left( \begin{array}{ccc}
        0 & -x_3 & x_2\\
        x_3 & 0 & -x_1\\
        -x_2 & x_1 & 0
\end{array}
\right)  \, v_u \; .
\end{eqnarray}
%
Apart from this type of contribution we could, in principle,
find contributions arising at the relative order $\epsilon$
due to the perturbation of the VEVs of the
flavons at this relative order, when NLO
corrections are included into the flavon (super-)potential.
However, the coupling from which the LO term in eq. (\ref{LOmd})
originates is generated at the renormalizable level, i.e. without
involving a flavon. Thus, the most general
NLO corrections to the neutrino Dirac mass matrix,
$\delta m_D$, are of the form given in eq. (\ref{deltamD}).
In explicit models the term $\delta m_D$
has usually a special form.
On the one hand,
the flavons in triplet representations have
a certain alignment, such as $(1,1,1)^t$, $(1,0,0)^t$, $(0,1,0)^t$
or $(0,0,1)^t$. On the other hand, in such models
usually there exist two different
flavour symmetry breaking sectors
which are separated by an additional
cyclic symmetry. In most cases
each of these sectors contains one triplet of flavons.
Considering NLO corrections arising at the level
of one flavon insertions, we expect that at most
fields from one of the two flavour symmetry
breaking sectors can couple at the
NLO level to give rise to corrections to
the neutrino Dirac mass matrix. Thus, there is only
one flavon triplet contributing to $\delta m_D$ at this level.
In the specific framework of the AF model, the
NLO terms are given by the triplet flavon $\varphi_T$
with $\langle \varphi_T \rangle = v_T \, \epsilon \, \La (1,0,0)^t$
($v_T$ is complex with $|v_T| \sim \mathcal{O}(1)$), so that we find:
\begin{equation}\label{deltamDAF}
\delta m_D = y_S^\nu \, v_T \, \epsilon \left( \begin{array}{ccc}
        2 & 0 & 0\\
        0 & 0 & -1\\
        0 & -1 & 0
\end{array}
\right) \, v_u +
y_A^\nu  \, v_T \, \epsilon \left(\begin{array}{ccc}
        0 & 0 & 0\\
        0 & 0 & -1\\
        0 & 1 & 0
\end{array}
\right) \, v_u \; .
\end{equation}
%
In contrast,  in the AM model we find
that the triplet $\varphi_S$
with $\langle \varphi_S \rangle = v_S \, \epsilon \, \La (1,1,1)^t$
gives rise to the NLO terms such that:
\footnote{A contribution from the flavon $\xi$ transforming as a
trivial singlet under $A_4$ can be absorbed into the LO result,
as we have already indicated.}
\begin{equation}\label{deltamDAM}
\delta m_D = y_S^\nu  \, v_S \, \epsilon \left( \begin{array}{ccc}
         2 & -1 & -1\\
        -1 & 2 & -1\\
        -1 & -1 & 2
\end{array}
\right) \, v_u +
y_A^\nu \,  v_S \, \epsilon \left(\begin{array}{ccc}
        0 & -1 & 1\\
        1 & 0 & -1\\
        -1 & 1 & 0
\end{array}
\right) v_u \; .
\end{equation}
%

 Similar to the neutrino Dirac mass matrix,
the matrices $m_M$ and $m_l$ also receive
corrections at the NLO level through multi-flavon insertions
and shifts in the flavon VEVs.
These corrections generate small off-diagonal
elements in the charged lepton mass matrix $m_l$.
If the corrections are of general type,
the matrix $V_{eL}$ satisfying:
\begin{equation}
V_{eL}^\dagger m_l^\dagger m_l \, V_{eL} =
\mbox{diag} (m_e^2, m_\mu^2, m_\tau^2)\,,
\end{equation}
%
has the form:
\begin{equation}
\label{VeL}
V_{eL} \approx \left( \begin{array}{ccc}
    1 & z_A \epsilon & z_B \epsilon\\
    -\overline{z_A} \epsilon & 1 & z_C \epsilon\\
    -\overline{z_B} \epsilon & -\overline{z_C} \epsilon & 1
\end{array}
\right)
\end{equation}
%
where $\overline{z}$ denotes the complex conjugate of $z$.
The parameters $z_i$ are, in general, complex numbers
and $|z_i| \sim \mathcal{O}(1)$.
The Majorana mass matrix $m_M$ of the RH neutrinos also
gets contributions from NLO corrections $\delta m_M$,
so that it is no longer exactly
diagonalized by $U_{TB}$, i.e. we have:
\begin{equation}
V_R^T \hat{U}_{TB}^T \left( m_M + \delta m_M \right) \hat{U}_{TB} V_R = \mbox{diag} (\tilde M_1, \tilde M_2, \tilde M_3)\,,
\end{equation}
where $V_R$ is defined by:
\begin{equation}
\label{VR}
V_R \approx \left( \begin{array}{ccc}
    1 & w_A \epsilon & w_B \epsilon\\
    -\overline{w_A} \epsilon & 1 & w_C \epsilon\\
    -\overline{w_B} \epsilon & -\overline{w_C} \epsilon & 1
\end{array}
\right) \; .
\end{equation}
%
The mass eigenvalues $\tilde M_i$ are expected
to differ from those calculated at LO, $M_i$,
by relative corrections of order
$\epsilon$. Also here the complex parameters $w_i$
have absolute values $|w_i| \sim \mathcal{O}(1)$.
We show in both matrices, $V_{eL}$ and $V_R$,
the leading term in the expansion
in $\epsilon$ for each matrix element.
In the two models we discuss in more detail
one finds \cite{AF2,AM} that due to the
structure of the NLO terms, not all parameters
$z_i$ and $w_i$ in $V_{eL}$ and $V_R$, respectively, are
arbitrary: in the AF model we have $z_A=z_B=z_C$
with no constraints on $w_i$, while in the AM model
$z_i$ are not related, but $w_A=0$ and $w_C=0$.

  Finally, we comment in a more quantitative
way on the size of the expansion parameter
$\epsilon$, which in turn entails
constraints on the possible size of
$\tan\beta= \langle h_u \rangle/\langle h_d \rangle=v_u/v_d$.
The upper bound on  $\epsilon$ comes from
the requirement that the discussed NLO corrections
to the lepton mixing angles do not lead to too large deviations
from the experimental best fit values.
The strongest constraint results from the
data on the solar neutrino mixing
angle and implies $\epsilon \lesssim 0.05$.
A lower bound on $\epsilon$ can be obtained by
taking into account the fact that the Yukawa coupling of the
$\tau$ lepton should not be too large. As mentioned,
a rather generic feature of the models of interest
is that the $\tau$ lepton mass is generated through a
non-renormalizable operator involving one flavon. As a consequence,
the following relation holds:
\begin{equation}
m_\tau \approx |y_\tau| \epsilon \langle h_d \rangle = |y_\tau| \epsilon \frac{v}{\sqrt{2}} \cos\beta
\approx |y_\tau| \epsilon \frac{v}{\sqrt{2}} \frac{1}{\tan\beta}
\end{equation}
%
where $v \approx 246$ GeV.
Taking $m_\tau$ at the $Z$ mass scale,  $m_\tau (M_Z) \approx 1.74$ GeV
\cite{mtauatMZ}, we find:
\begin{equation}
0.01 \approx |y_\tau| \frac{\epsilon}{\tan\beta} \; .
\end{equation}
%
Reasonable values for $|y_\tau|$ are between $1/3$ and $3$.
Using $|y_\tau|=3$ and $\tan\beta=2$ gives:
\footnote{As is well known, $\tan\beta$ cannot be too small
\cite{tanbetabound}. We allow here for the rather
low value of $\tan\beta = 2$.}
\begin{equation}
\epsilon\approx 0.007 \; .
\end{equation}
%
This is the minimal value of
$\epsilon$ in this type of models.
For $\epsilon \approx 0.05$
one finds that $|y_\tau|=3$ corresponds to the
largest allowed value of $\tan\beta=15$.
All smaller values of $\tan\beta \gtrsim 2$
are possible as well.
In the numerical analysis we fix
$\epsilon =0.04$. In this case the corresponding
allowed range of $\tan\beta$ is
$2 \lesssim \tan\beta \lesssim 12$.

\cleqn
\section{Charged Lepton Flavour Violating Radiative Decays}

\subsection{Basic Formulae}

  We calculate the branching ratios of the LFV processes
$\ell_i\rightarrow \ell_j+\gamma$ ($m_{\ell_i}>m_{\ell_j}$) using
the following expression \cite{BR1,BR2}:
\begin{equation}
B(\ell_i\rightarrow \ell_j+\gamma)\;\approx\;B(\ell_i\rightarrow
\ell_j+\nu_i+\bar{\nu}_j)\, B_0(m_0,m_{1/2})\,\left|\sum_{k}
(\hat{Y}_\nu^\dagger)_{ik}\log\left(\frac{M_X}{M_k}\right)(\hat{Y}_\nu)_{kj}
\right|^2\tan^2\beta\label{BR}
\end{equation}
%
where $\ell_1= e$, $\ell_2=\mu$ and $\ell_3=\tau$.
In eq. (\ref{BR}) $\hat{Y}_\nu$ is the matrix
of neutrino Yukawa couplings, computed taking into account
all NLO effects in the basis in which the
charged lepton and RH neutrino mass
matrices are diagonal and have positive eigenvalues:
\begin{equation}
 \hat{Y}_\nu \;=\; V_R^T\,\Omega\,U_{TB}^T\,Y_\nu\,V_{eL}\label{Ynu}
\end{equation}
%
where $\Omega$ is introduced in eq. (\ref{UTB})
and $Y_\nu$ represents the matrix of neutrino Yukawa couplings in the
basis in which the superpotential is defined (see eq.
(\ref{YnuOr})). We consider the neutrino Dirac mass
matrix $m_D$, including the generic NLO corrections given in
eq. (\ref{deltamD}). The unitary matrices $V_R$ and
$V_{eL}$ are given in eqs. (\ref{VR}) and (\ref{VeL}),
respectively.

 According to the mSUGRA scenario we consider,
at the scale of grand unification $M_X\approx 2\times 10^{16}$
GeV, the slepton mass matrices are diagonal and
universal in flavour and the trilinear couplings are proportional
to the Yukawa couplings:
\begin{eqnarray}
&&
(m^2_{\tilde{L}})_{ij}=(m^2_{\tilde{e}})_{ij}=(m^2_{\tilde{\nu}})_{ij}=\delta_{ij}
m^2_0 \; ,\\
&&(A_{\nu})_{ij}=A_0(Y_{\nu})_{ij} \; ,\\
&&(A_{e})_{ij}=A_0(Y_{e})_{ij}\;,~~A_0 = a_0 m_0\,,
\end{eqnarray}
%
where $m_{\tilde{L}}^2$ and $m_{\tilde{e}}^2$ are the left-handed and
right-handed charged slepton mass matrices, respectively, while
$m_{\tilde{\nu}}^2$ is the right-handed sneutrino soft mass term.
The gaugino masses are assumed to have a common
value at the high scale $M_X$:
\begin{equation}
M_{\widetilde{B}}=M_{\widetilde{W}}=M_{\widetilde{g}}=m_{1/2} \; .
\end{equation}
%
The scaling function $B_0(m_0,m_{1/2})$ contains
the dependence on the SUSY breaking parameters:
\begin{equation}
B_0(m_0,m_{1/2})\;\approx\;\frac{\alpha_{em}^3}{G_F^2 m_S^8
}\left|\frac{(3+a_0^2)m_0^2}{8\pi^2} \right|^2 \; .\label{B0}
\end{equation}
%
In eq. (\ref{B0}), $G_F$ is the Fermi constant and
$\alpha_{em}\approx 1/137$ is the fine structure
constant.
The SUSY mass parameter $m_S$ in eq. (\ref{B0})
was obtained by performing a fit to
the exact RG results \cite{BR1}.
The resulting analytic
expression in terms of $m_0$ and $m_{1/2}$
has the form \cite{BR1}:
\begin{equation}
 m_S^8\approx 0.5\, m_0^2\, m_{1/2}^2\,(m_0^2+0.6\, m_{1/2}^2)^2 \; .\label{mS}
\end{equation}
%
According to \cite{BR1}, deviations from the exact RG
result can be present in the region
of relatively large (small) $m_{1/2}$
and small (large) $m_0$.
\begin{figure}[t!!]
\begin{center}
\includegraphics[width=10.5cm,height=7.5cm]{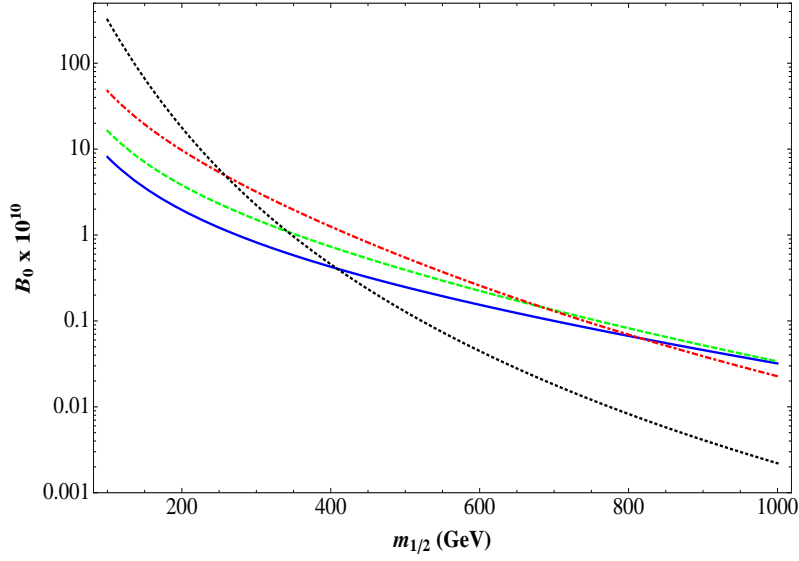}
\caption{The dependence of the scaling function $B_0(m_0,m_{1/2})$
(see eq. (\ref{B0})) on $m_{1/2}$ for $A_0=0$ and fixed $m_0$: $i)$ $m_0=100$
GeV (black, dotted line), $ii)$ $m_0=400$ GeV (red, dot-dashed
line), $iii)$ $m_0=700$ GeV (green, dashed line) and $iv)$
$m_0=1000$ GeV (blue, continuous line).
\label{Fig1}}
\end{center}
\end{figure}
%
In Fig. \ref{Fig1} we show the dependence of $B_0(m_0,m_{1/2})$
on $m_{1/2}$ for fixed values of $m_0$ and $A_0=0$.
We notice that the function $B_0$ and, consequently,
the branching ratio in eq. (\ref{BR}), can vary up to three to four
orders of magnitude, depending on which point ($m_0$,
$m_{1/2}$) of the parameter space is considered.
For smaller values of $m_{1/2}$ we have larger LFV branching ratios or,
equivalently, the larger is the SUSY mass parameter $m_S$,
the stronger is the suppression of the predicted branching ratio.
Values of $A_0\neq 0$ lead to larger values of $B_0 (m_0, m_{1/2})$
(see eq. (\ref{B0})) and to an increase of
the branching ratios, see eq. (\ref{BR}). In the numerical analysis
we set $A_0=0$, unless
otherwise specified.

%
\mathversion{bold}
\subsection{Identifying the Leading Order Contributions in $B(\ell_i\rightarrow \ell_j+\gamma)$}
\mathversion{normal}
%
%

 We work for convenience with LFV branching ratios
given in eq. (\ref{BR}), normalized to the partial branching ratios of
the $\mu$ or $\tau$ decays into lighter charged lepton and two neutrinos:
\begin{equation}
 B'(\ell_i\rightarrow \ell_j+\gamma)\;=\;\frac{B(\ell_i\rightarrow \ell_j+\gamma)}{B(\ell_i\rightarrow \ell_j+\nu_i+\bar{\nu}_j)} \; .
\label{BRprime}
\end{equation}
%
We have: $B(\mu\rightarrow e+\gamma)\approx B'(\mu\rightarrow
e+\gamma)$, $B(\tau\rightarrow e+\gamma)\approx 0.18\,
B'(\tau\rightarrow e+\gamma)$ and
$B(\tau\rightarrow\mu+\gamma)\approx 0.17\,
B'(\tau\rightarrow\mu+\gamma)$ \cite{Amsler:2008zzb}.
\begin{table}
\begin{center}
\begin{tabular}{|c|c|}
\hline
$\mu\;\rightarrow\;e\;+\;\gamma$ & \\
\hline $\hspace{0.5cm}(\hat{Y}_\nu^\dagger
\hat{Y}_\nu)_{21}\hspace{0.5cm}$ & \hspace{1cm}
$y_\nu\left(\,w^{\prime\prime} \overline{y}^\nu_{1'} + w^\prime y^\nu_{1''}-x_2 (y^\nu_A+y^\nu_S) -x_3 (\overline{y}^\nu_A+\overline{y}^\nu_S)\,\right)\e+\mathcal{O}(\e^2)\hspace{1cm}$\\
$\hspace{0.5cm}(\hat{Y}_\nu^\dagger)_{22}(\hat{Y}_\nu)_{21}\hspace{0.5cm}$
& \hspace{1cm}
$\frac{1}{3}\,y_\nu^2\;+\;\mathcal{O}(\e)\hspace{1cm}$\\
$\hspace{0.5cm}(\hat{Y}_\nu^\dagger)_{23}(\hat{Y}_\nu)_{31}\hspace{0.5cm}$
& \hspace{1cm} $\mathcal{O}(\e)\hspace{1cm}$ \\ \hline
$\tau\;\rightarrow\;e\;+\;\gamma$ & \\
\hline $\hspace{0.5cm}(\hat{Y}_\nu^\dagger
\hat{Y}_\nu)_{31}\hspace{0.5cm}$ & \hspace{1cm}
$y_\nu\left(\,w^{\prime\prime}y^\nu_{1'} + w^\prime\overline{y}^\nu_{1''}+ x_2 (\overline{y}^\nu_A-\overline{y}^\nu_S) + x_3 (y^\nu_A-y^\nu_S) \,\right)\e+\mathcal{O}(\e^2)\hspace{1cm}$\\
$\hspace{0.5cm}(\hat{Y}_\nu^\dagger)_{32}(\hat{Y}_\nu)_{21}\hspace{0.5cm}$
& \hspace{1cm}
$\frac{1}{3}\,y_\nu^2\;+\;\mathcal{O}(\e)\hspace{1cm}$\\
$\hspace{0.5cm}(\hat{Y}_\nu^\dagger)_{33}(\hat{Y}_\nu)_{31}\hspace{0.5cm}$
& \hspace{1cm}
$\mathcal{O}(\e)\hspace{1cm}$\\
\hline
$\tau\;\rightarrow\;\mu\;+\;\gamma$ & \\
\hline $\hspace{0.5cm}(\hat{Y}_\nu^\dagger
\hat{Y}_\nu)_{32}\hspace{0.5cm}$ & \hspace{1cm}
$y_\nu\left(\,w^{\prime\prime}\overline{y}^\nu_{1'} + w^\prime y^\nu_{1''}+2\, x_2\, y^\nu_S +2\,x_3\, \overline{y}^\nu_S\,\right)\e+\mathcal{O}(\e^2)\hspace{1cm}$\\
$\hspace{0.5cm}(\hat{Y}_\nu^\dagger)_{32}(\hat{Y}_\nu)_{22}\hspace{0.5cm}$
& \hspace{1cm}
$\frac{1}{3}\,y_\nu^2\;+\;\mathcal{O}(\e)\hspace{2cm}$\\
$\hspace{1cm}(\hat{Y}_\nu^\dagger)_{33}(\hat{Y}_\nu)_{32}\hspace{1cm}$
& \hspace{1cm}
$-\frac{1}{2}\,y_\nu^2\;+\;\mathcal{O}(\e)\hspace{1cm}$\\
\hline
\end{tabular}
\end{center}
\begin{center}
\caption{Combination of elements of the matrix of neutrino Yukawa
couplings, $\hat{Y}_\nu$, which enter into the expression for
the branching ratios of the LFV
decay $\ell_i\rightarrow \ell_j +\gamma$
(see eq. (\ref{BR_logs})). The expression
for the relevant $\mathcal{O}(\e)$
terms in $(\hat{Y}^\dagger_\nu\hat{Y}_\nu)_{ij}$ ($i\neq j$) is
also given (see text for details).
\label{tab}}
\end{center}
\end{table}
%

  It proves useful to analyze separately
the contributions in the LFV branching ratios,
which are associated with each of the three heavy
RH Majorana neutrinos.
For this purpose we rearrange the
terms in eq. (\ref{BR}) in the following way:
\begin{equation}
B'(\ell_i\rightarrow \ell_j+\gamma)\propto
\left|(\hat{Y}_\nu^\dagger\hat{Y}_\nu)_{ij}\log\left(\frac{m_1}{m_*}\right)+
(\hat{Y}_\nu^\dagger)_{i2}(\hat{Y}_\nu)_{2j}\log\left(\frac{m_2}{m_1}\right)+
(\hat{Y}_\nu^\dagger)_{i3}(\hat{Y}_\nu)_{3j}\log\left(\frac{m_3}{m_1}\right)\right|^2\label{BR_logs}
\end{equation}
%
with
\begin{equation}
 m_*\;=\; \frac{v_u^2 y_\n^2}{M_X}\cong (1.5\times 10^{-3}\,{\rm
eV})\,y_\nu^2\,\sin^2\beta\approx 1.5\times 10^{-3}\,{\rm
eV}\label{mstar}
\end{equation}
%
where we have fixed $y_\nu=1$ and
used $\sin^2\beta\approx 1$,
which is a good approximation
given the fact that $\tan\beta\gtrsim 2$.
In eq. (\ref{BR_logs}), $m_k$, $k=1,2,3$, are
the LO neutrino masses in the $A_4$ models,
given by eq. (\ref{numass}) and generated via the
type I see-saw mechanism.
We neglect contributions to $m_k$ which arise
from NLO corrections in the superpotential
since the branching ratio depends only logarithmically
on the light neutrino masses and typically
all such relative corrections are of order
$\e\approx(0.007\div0.05)$,
as discussed in the previous section.
We also do not consider RG effects
in the calculation of the neutrino masses and mixings.
Such corrections can be relevant in the case of
quasi-degenerate (QD) light neutrino mass spectrum. These
corrections are relatively small or  negligible
if the spectrum is hierarchical,
or with partial hierarchy \cite{RGnuref}. In the $A_4$ models under
discussion, the lightest neutrino mass in the case
of  NO (IO) mass spectrum
is constrained (by the data on the neutrino oscillation
parameters, see, e.g. \cite{AM,HMP09})
to lie in the interval
$3.8\times 10^{-3}\,{\rm eV}\lesssim m_1\lesssim 7\times 10^{-3}\,{\rm eV}$
($0.02\,{\rm eV}\lesssim m_3$). Thus, in the case of IO
spectrum we present results for
$0.02\,{\rm eV}\lesssim m_3 \lesssim 0.10\,{\rm eV}$.
\begin{figure}[t!!]
\begin{center}
\begin{tabular}{cc}
\includegraphics[width=7.5cm,height=6.5cm]{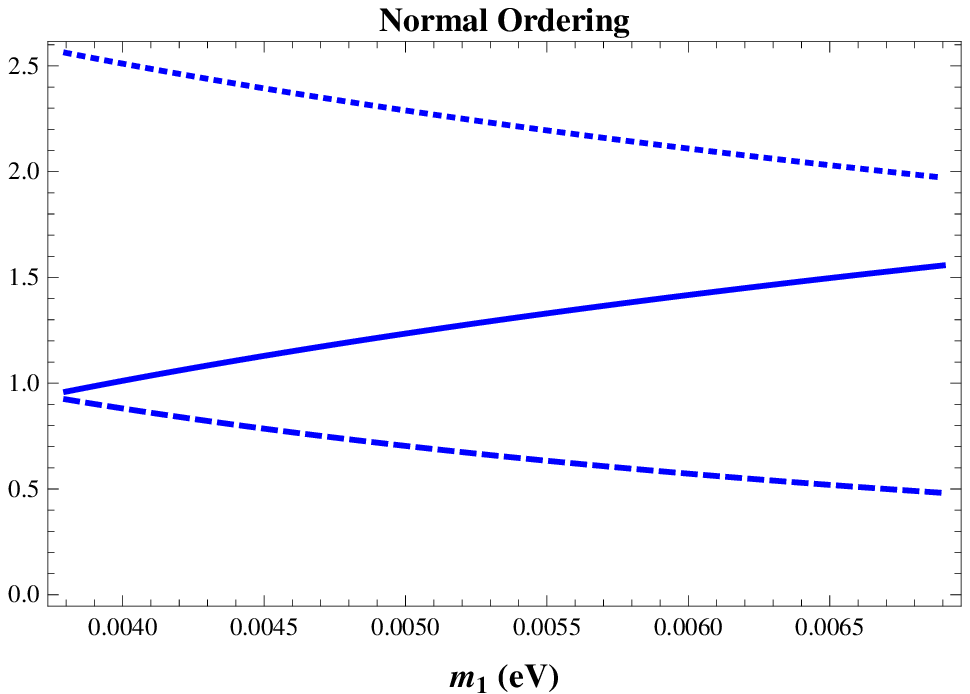} &
\includegraphics[width=7.5cm,height=6.5cm]{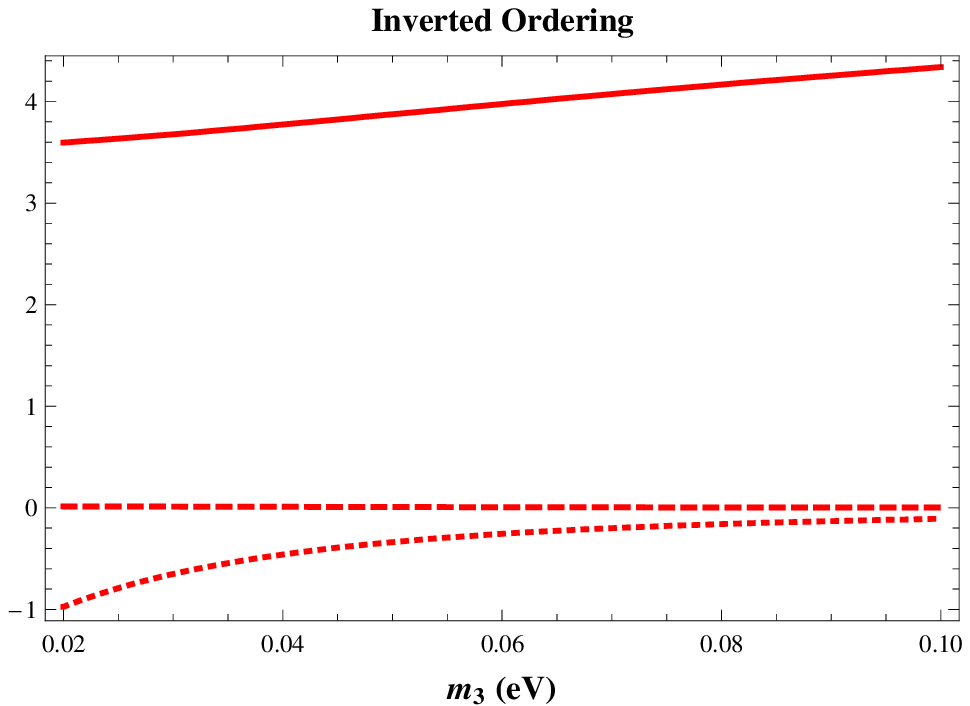}
\end{tabular}
\caption{The three different contributions in
$B'(\ell_i\rightarrow \ell_j+\gamma)$,
eq. (\ref{BR_logs}), in the case of light neutrino mass spectrum with
normal (inverted) ordering (left (right) panel):
$i)$ $\log\left(m_1/m_*\right)$ vs $m_1~ (m_3)$ (continuous line), $ii)$
$\log\left(m_2/m_1\right)$ vs $m_1~(m_3)$ (dashed line) and $iii)$
$\log\left(m_3/m_1\right)$ vs $m_1~(m_3)$ (dotted line).
The results shown correspond to
the best fit values \cite{nudata}
$|\dma|=2.40\times10^{-3}$ eV$^2$ and
$r=\dmsol/|\dma|=0.032$.
\label{Fig2}}
\end{center}
\end{figure}
%
As we will see,
the predictions for the branching ratios of the LFV
decays $\ell_i\rightarrow \ell_j+\gamma$
depend, in general, on the type of neutrino mass spectrum.

\paragraph{Neutrino Mass Spectrum with Normal Ordering\\\\}

In the case of NO mass spectrum, the three logarithms in
eq. (\ref{BR_logs}) are all positive and are of the same order
(see Fig. \ref{Fig2}, left panel).
The dominant contribution to the decay amplitude depends
strongly on the combination of
$\hat{Y}_\nu$ matrix elements in eq. (\ref{BR_logs}).
Note that the matrix elements of
$\hat{Y}_\nu$ take all $\mathcal{O}(1)$
values, except for the $(31)$ entry
which  typically scales as the
expansion parameter $\e$. This is due to the presence of the
TB mixing matrix $U_{TB}$ in the expression for the neutrino
Yukawa couplings $\hat{Y}_\nu$, eq.  (\ref{Ynu}).

 In Table \ref{tab} we give the order of magnitude
in $\epsilon$ of the coefficients of the three logarithms,
$(\hat{Y}_\nu^\dagger\hat{Y}_\nu)_{ij}$,
$(\hat{Y}_\nu^\dagger)_{i2}(\hat{Y}_\nu)_{2j}$ and
$(\hat{Y}_\nu^\dagger)_{i3}(\hat{Y}_\nu)_{3j}$,
which appear in the three branching ratios
$B'(\ell_i\rightarrow \ell_j+\gamma)$
of  interest. The VEVs of the flavon fields are
assumed to be real for simplicity.
As Table \ref{tab} shows,
the coefficient of the $\log(m_2/m_1)$ term
in each of the three LFV branching ratios
under discussion is  of order one.
The same conclusion is valid for the coefficient of
the $\log(m_3/m_1)$ term
in $B'(\tau\rightarrow\mu+\gamma)$.
In what concerns the coefficients
of the term $\propto \log(m_1/m_*)$,
they always originate from NLO
corrections in the superpotential.
These coefficients correspond
to the off-diagonal elements of the hermitian matrix
$\hat{Y}_{\nu}^\dagger \hat{Y}_{\nu}$,
in which the rotation
matrices, $U_{TB}$, $\Omega$ and  $V_R$, associated with the
diagonalisation of the RH neutrino
Majorana mass term, do not appear. At order
$\e$, they depend only on the parameters
of the neutrino Dirac  mass matrix
(see eqs. (\ref{mD}) and (\ref{deltamD})),
as reported in Table \ref{tab}.

  Taking into account the magnitude of the
different terms
shown in Table \ref{tab}, we
expect, in general, that in the case of NO neutrino
mass spectrum:
\begin{eqnarray}
B^\prime(\mu\rightarrow e +\gamma)\approx
B^\prime(\tau\rightarrow e
+\gamma)\approx B_0(m_0,m_{1/2})\left|\frac{1}{3}y_\nu^2\log\left(\frac{m_2}{m_1}\right)\right|^2\tan^2\beta\propto
0.1\left|y_\nu\right|^4 \; , \label{BRNOest1}\\
B^\prime(\tau\rightarrow
\mu+\gamma)\approx B_0(m_0,m_{1/2})\left|\frac{1}{3}y_\nu^2\log\left(\frac{m_2}{m_1}\right)-\frac{1}{2}y_\nu^2\log\left(\frac{m_3}{m_1}\right)\right|^2\tan^2\beta
\propto\left|y_\nu\right|^4\;. \label{BRNOest2}
\end{eqnarray}
%

From the analytical estimates,
eqs. (\ref{BRNOest1}) and (\ref{BRNOest2}),
we conclude that in the case of NO mass spectrum,
$B^\prime(\tau\rightarrow \mu+\gamma)$ is
approximately by one order of magnitude larger
than $B^\prime(\tau\rightarrow e+\gamma)$ and
$B^\prime(\mu\rightarrow
 e+\gamma)$.

\paragraph{Neutrino Mass Spectrum with Inverted Ordering\\\\}

 As can be seen in Fig. \ref{Fig2} (right panel),
the term proportional to $\log\left(m_2/m_1\right)$
 is strongly suppressed with respect to
the other terms in the case of IO mass spectrum. This is valid for
all values of the lightest neutrino mass allowed in the models of
interest, $m_3\gtrsim 0.02$ eV. More specifically, one has:
$\log\left(m_2/m_1\right)\approx 0.014$ (0.003) for $m_3=0.02$ eV
($0.1$ eV). In the case of non-QD neutrino mass spectrum
($m_3\lesssim 0.1$ eV), $B'(\ell_i\rightarrow \ell_j+\gamma)$ are
determined practically by the sum of the terms proportional to
$\log(m_1/m_*)$ and $\log(m_3/m_1)$. The second term increases as
$m_3$ decreases towards the minimal allowed value $m_3 \approx
0.02$ eV, so that $|\log(m_3/m_1)|\approx 1$ ($0.1$)
for $m_3=0.02$ eV ($0.1$ eV).
Thus, taking into account
the results reported in Table \ref{tab},
we have in the LO approximation in $\epsilon$:
\begin{equation}
B^\prime(\mu\rightarrow e +\gamma) \;\approx\;
B^\prime(\tau\rightarrow e +\gamma)\;\propto\;\,\mathcal{O}(\e^2)
\label{BRIOest1} \; ,
\end{equation}
\begin{equation}
 B^\prime(\tau\rightarrow \mu+\gamma)\;
\propto \;
\left|\frac{1}{2}y_\nu^2\log\left(\frac{m_3}{m_1}\right)\right|^2
\approx\left\{
\begin{array}{c}
0.25\,|y_\nu|^4\,,\,\,\,{\rm for\,}m_3=0.02\,{\rm eV}\;,\\
0.0025\,|y_\nu|^4\,,\,\,\,{\rm for\,}m_3=0.1\,{\rm eV}\;.
\end{array}\right.
\label{BRIOest2} 
\end{equation}
%

The shown
order of magnitude estimates for $B^\prime(\mu\to e
+\gamma)$ and $B^\prime(\tau\to e+\gamma)$ in eq. 
(\ref{BRIOest1}) can be significantly modified by the rather
large contribution of the term containing the factor
$\log(m_1/m_*)\approx (3.5\div 4.5)$. It follows from
Table \ref{tab} that for, e.g. $\e \approx 0.04$, the contribution
in the LFV branching ratios due to the indicated term can be $\sim
\e\,\log(m_1/m_*) \approx 1/5\sim \sqrt{\epsilon}$ such that the branching ratios of the decays $\mu\to  e +
\gamma$ and $\tau\to  e +\gamma$  scale as $\mathcal{O}(\epsilon)$.
For $m_3\approx0.1$ eV, $B^\prime(\mu\to e
+\gamma)$ and $B^\prime(\tau\to e+\gamma)$ can be comparable to the normalized
branching ratio of $\tau\to\mu + \gamma$ decay,
eq. (\ref{BRIOest2}). Indeed, for $m_3\approx 0.1$ eV and
$\epsilon\approx 0.04$, owing to the interplay between the
leading term in the expansion parameter $\epsilon$,
$\log(m_3/m_1)$, whose absolute value decreases with increasing of $m_3$, and the contribution from
$\log(m_1/m_*)$, $B^\prime(\tau\rightarrow \mu+\gamma)$ scales as
few times $\epsilon$.

Comparing the results for the NO and the IO neutrino mass spectrum
we see that in a model with generic NLO corrections to the matrix
of neutrino Yukawa couplings, the magnitude of the branching ratio
$B^\prime(\tau\rightarrow \mu+\gamma)$ practically does not depend
on the type of neutrino mass spectrum. For $\e \approx 0.007$,
$B^\prime(\mu\rightarrow e +\gamma)$ and $B^\prime(\tau\rightarrow
e+\gamma)$ in the case of IO spectrum can be by one order of
magnitude smaller than in the case of NO spectrum, while if $\e
\approx 0.04$, these two branching ratios are predicted to be
essentially the same for the two types of spectrum. We always have
(independently of the type of the spectrum and of the value of
$\e$) $B^\prime(\mu\rightarrow e +\gamma) \approx
B^\prime(\tau\rightarrow e+\gamma)$.

   In the next section we study numerically the LFV processes in the
AF and AM models. One important difference between the two models is
in the predicted off-diagonal elements of the hermitian matrix
$\hat{Y}_\nu^\dagger\hat{Y}_\nu$. In the AM model they are all of
$\mathcal{O}(\e)$ and originate from the NLO corrections
to the Dirac mass matrix, eq.  (\ref{deltamDAM}).
The exact expressions for the matrix elements can
be derived  using  Table \ref{tab} and setting
$w^\prime=w^{\prime\prime}=0$ and $(x_1,x_2,x_3)=(1,1,1) \, v_S$.
In what concerns the AF model, the VEV structure of the
flavon fields, $w^\prime=w^{\prime\prime}=0$ and $(x_1,x_2,x_3)\propto(1,0,0)$,
implies that the leading term in the off-diagonal elements of the matrix
$\hat{Y}_\nu^\dagger\hat{Y}_\nu$ is of $\mathcal{O}(\e^2)$.
This receives contributions from the Dirac mass
term, see eq. (\ref{deltamDAF}), as well as from the
charged lepton sector (through $V_{eL}$, see eq. (\ref{VeL})).
This difference in
the $\epsilon$ dependence of the elements of
$\hat{Y}_\nu^\dagger\hat{Y}_\nu$ in the two models
leads to different predictions
for the LFV branching ratios for the IO neutrino mass spectrum.
As a consequence, in the AM model
the branching ratios of the decays $\mu \to e +\gamma$ and $\tau\to e+\gamma$
are up to two orders of magnitude larger than those in the AF model.
In contrast, for $m_3< 0.1$ eV, we expect similar results
in both models for the decay $\tau\to\mu+\gamma$
since the coefficient
of the term proportional to $\log(m_3/m_1)$
is of order $\epsilon^0$.

  We note that in the case of a
QD light (heavy) neutrino mass spectrum,
$m_3 \gtrsim 0.1$ eV, 
the term proportional to
$\log(m_1/m_*)$ in eq. (\ref{BR_logs}) gives the dominant
contribution
and thus the magnitude of the non-diagonal
elements of $\hat{Y}_\nu^\dagger\hat{Y}_\nu$ determines
the magnitude of the
branching ratios of the LFV decays.

\vspace{0.1in}
The preceding discussion shows that
in the $A_4$ models, the LO structure of the
matrix of neutrino Yukawa couplings $\hat{Y}_\nu$, which is
determined by $U_{TB}$, together with the possibility
of having a heavy RH neutrino mass spectrum with
partial hierarchy, leads to
LFV decay rates scaling as $\mathcal{O}(\e^0)$.
This prediction differs significantly from
the one obtained in the effective field theory approach.
In \cite{EFTA4} the branching ratios of the
charged LFV radiative decays were
shown to scale as $\e^2$
in a generic effective field theory framework,
and could even be stronger suppressed (scaling as $\e^4$)
in a specific supersymmetric scenario.

 Concerning the absolute magnitude of the branching ratios
we remark that these are expected to be of similar size in both approaches,
because the suppression
due to (positive) powers of $\epsilon$ present
in the effective field theory approach corresponds in our
case to the suppression factor associated to the fact that
flavour violating soft slepton masses
are generated only through RG running.
The scales $m_S$ and $M$, which are the
relevant scales for charged LFV radiative decays,
in our approach and in the effective
field theory one \cite{EFTA4}, respectively,
can be related to each other. Assuming that the mass scale
$M$ arises from one-loop effects of new
particles, such as SUSY particles, we see that the mass $m_S$
of these new particles is identified with $M$
weighted with the coupling $g$
of these particles to the charged leptons and
divided by the loop factor $4\pi$.
 Thus, we roughly have:
$ m_S \sim g M/(4 \pi)$.

\cleqn
\section{Numerical Results}

   In this section we report results of the calculations
of branching ratios of the LFV decays
$\mu\rightarrow e+\gamma$, $\tau\rightarrow e+\gamma$ and
$\tau\rightarrow \mu+\gamma$.
This is done in the form of scatter plots
showing the correlations
between each two of the indicated branching ratios.
The calculations are performed
in the framework of the AF and AM models.
The expansion parameter $\e$ is set equal to 0.04
in the numerical analyses.

 We consider a scenario in which the sparticle
mass spectrum is moderately heavy:
\begin{eqnarray}
 && m_0=150\,\GeV,\,\,\,\,m_{1/2}=700\,\GeV,\,\,\,\,A_0=0\,\GeV,\,\,\,\,\tan\beta=10.\label{set}
\end{eqnarray}
%
The parameters in eq. (\ref{set}) lead to squark masses
between $1.1$ TeV and $1.5$ TeV, gluino
masses around $1.6$ TeV, and masses of right-handed
sleptons are $300$ GeV. Thus, these
sparticles are accessible at LHC.
This point in the mSUGRA parameter space
belongs to the stau co-annihilation region
\cite{Baer:2003wx,Ellis:2003cw,Baer:2009vr},
in which the amount of DM in the Universe
can be explained through
the lightest sparticle (LSP).
The latter is a bino-like neutralino
and has a mass of approximately $280$ GeV
\footnote{The sparticle masses quoted above
have been calculated with ISAJET 7.69 \cite{isajet}.}.
As has been shown in \cite{mSUGRAnuR},
the stau co-annihilation and the bulk regions
are hardly affected, if RH neutrinos are included into the mSUGRA context.
 For the set of parameters in eq. (\ref{set}), all decay rates scale
with the factor $B_0(m_0,m_{1/2})\tan^2\beta\approx 3.8\times10^{-10}$.

 The scatter plots are obtained by
varying all the $\mathcal{O}(1)$
parameters that enter in the matrix of neutrino Yukawa
couplings $\hat{Y}_\nu$, defined in eq. (\ref{Ynu}). Some of
these parameters are equal to zero or have a common value. More
specifically, in the AM model $w_A=w_C=0$ and in the AF model
$z_A=z_B=z_C$. The NLO corrections to the Dirac mass matrices
for the AF and AM models are given in eqs. (\ref{deltamDAF}) and
(\ref{deltamDAM}), respectively.
In the calculations of the normalized branching ratios
$B'(\ell_i\rightarrow \ell_j+\gamma)$, we set $y_\nu=1$ and the
absolute values of all the other (complex) parameters in
$\hat{Y}_\nu$ are varied in the interval $[0.5,2]$. The
corresponding phases are varied between 0 and $2\pi$.

\begin{figure}[t!!]
\begin{center}
\begin{tabular}{cc}
\includegraphics[width=7.5cm,height=6.5cm]{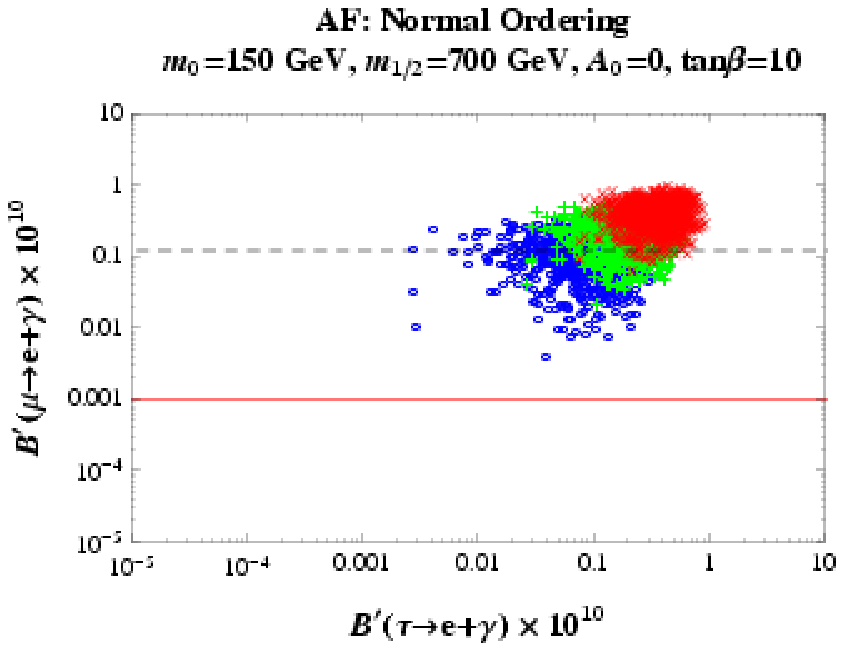} &
\includegraphics[width=7.5cm,height=6.5cm]{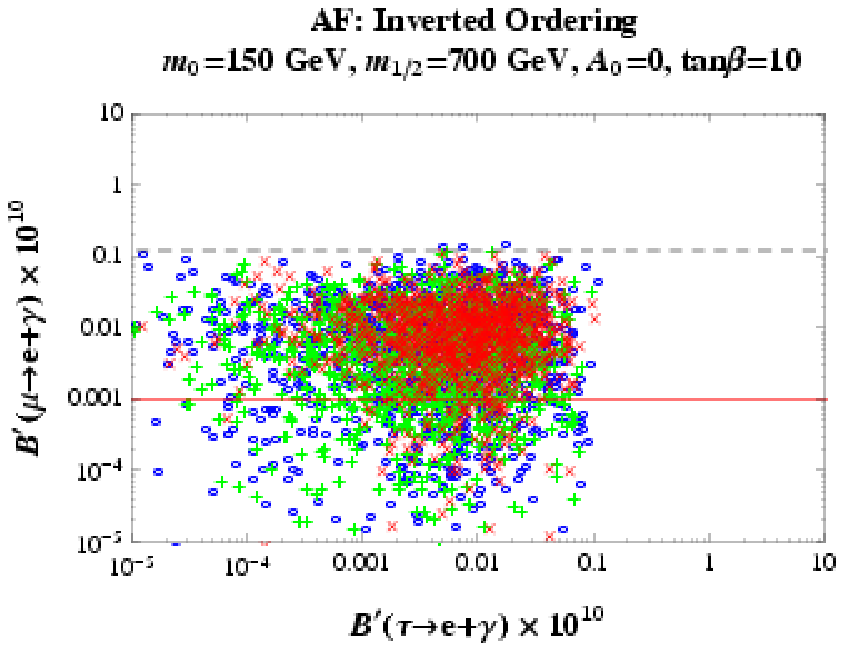}\\
\includegraphics[width=7.5cm,height=6.5cm]{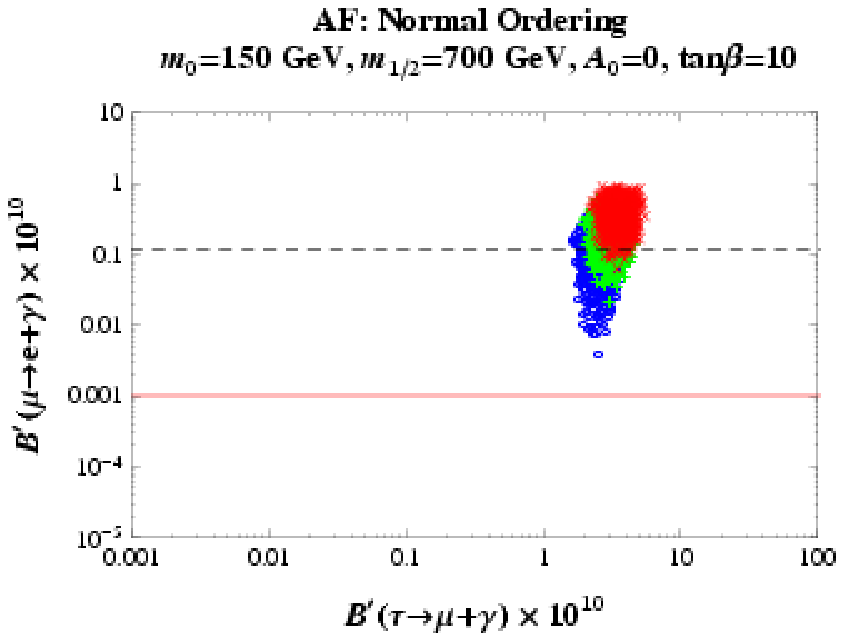} &
\includegraphics[width=7.5cm,height=6.5cm]{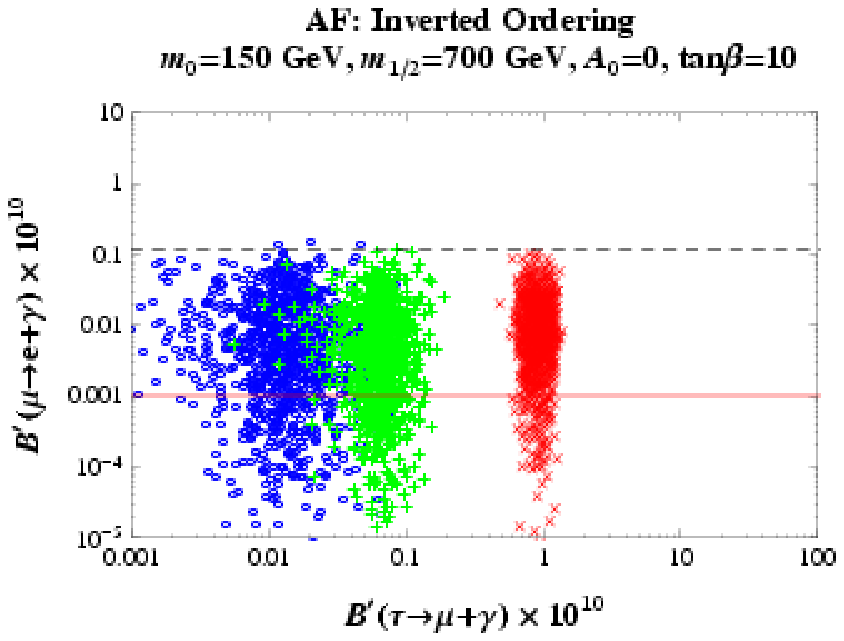}
\end{tabular}
\caption{
Correlation between
$B'(\mu\rightarrow e+\gamma)$,
$B'(\tau\rightarrow e+\gamma)$
and $B'(\tau\rightarrow \mu+\gamma)$, calculated in the AF model.
The results shown are obtained
for three different values of the
lightest neutrino mass for both types
of neutrino mass spectrum: $i)$ with normal
ordering (left panels), $m_1=3.8\times 10^{-3}$ eV (red $\times$),
$m_1=5\times 10^{-3}$ eV (green $+$) and $m_1=7\times 10^{-3}$ eV
(blue $\circ$); $ii)$ with inverted ordering (right panels), $m_3=0.02$
eV (red $\times$), $m_3=0.06$ eV (green $+$) and  $m_3=0.1$ eV
(blue $\circ$). The horizontal dashed line corresponds to the MEGA
bound \cite{MEGA}, $B'(\mu\rightarrow e +\gamma)\leq
1.2 \times 10^{-11}$. The horizontal continuous line
corresponds to $B'(\mu\rightarrow e +\gamma) =
10^{-13}$, which is the prospective sensitivity
of the MEG experiment \cite{MEG}.
\label{Fig3}}
\end{center}
\end{figure}

\begin{figure}[t!!]
\begin{center}
\begin{tabular}{cc}
\includegraphics[width=7.5cm,height=6.5cm]{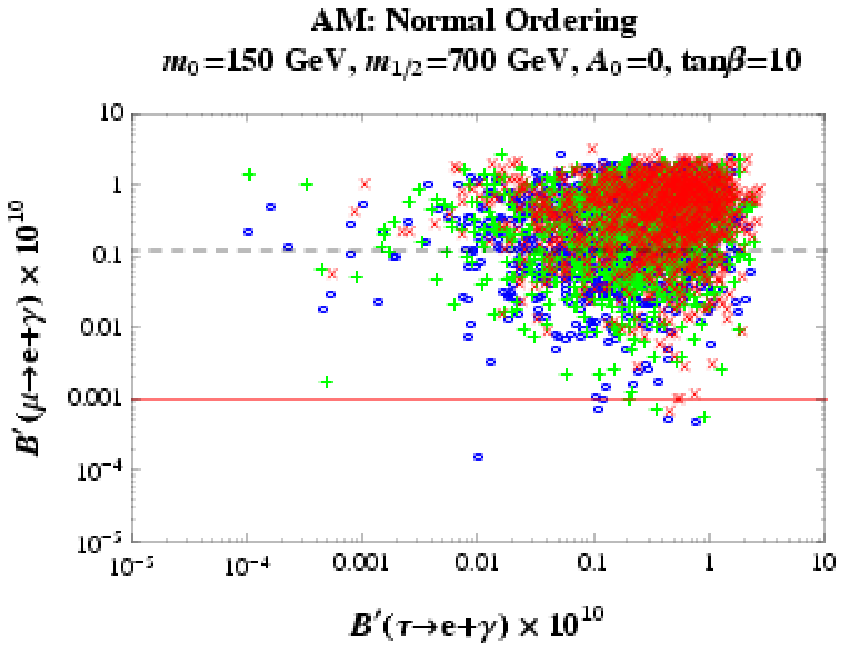} &
\includegraphics[width=7.5cm,height=6.5cm]{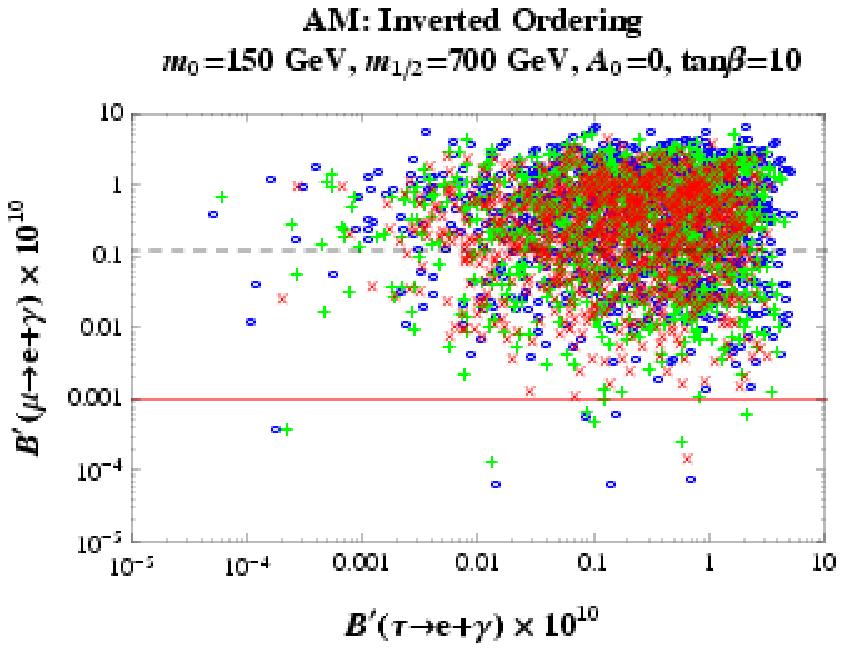}\\
\includegraphics[width=7.5cm,height=6.5cm]{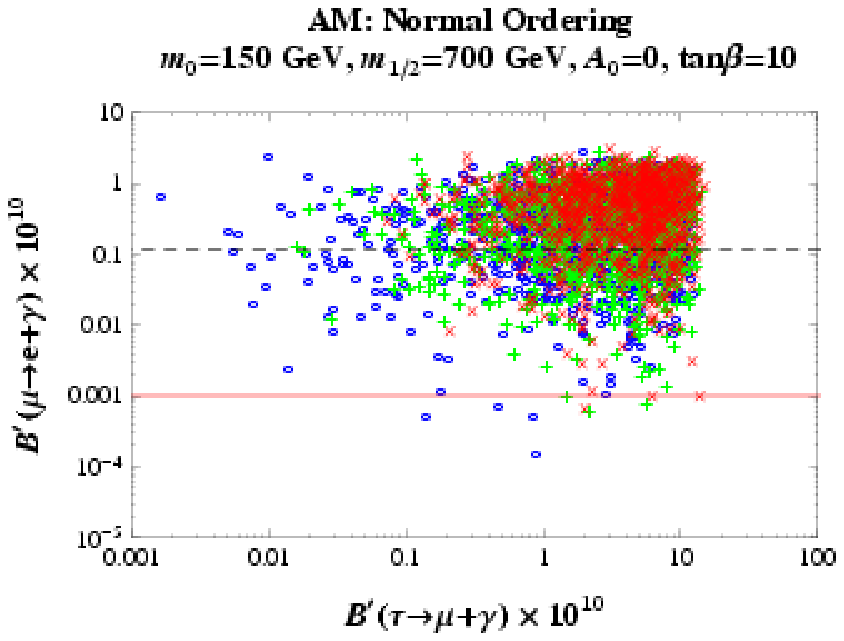} &
\includegraphics[width=7.5cm,height=6.5cm]{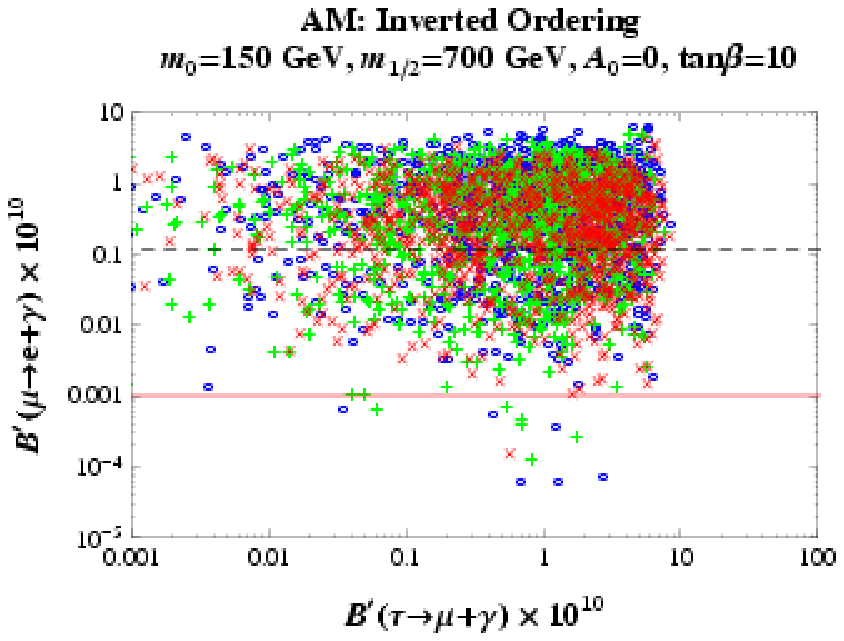}
\end{tabular}
\caption{The same as in Fig. \ref{Fig3}, but for the
AM model.
\label{Fig4}}
\end{center}
\end{figure}

  The results obtained for the AF and the AM models
and for both the NO and
IO light neutrino mass spectrum are presented graphically
in Figs. \ref{Fig3} and \ref{Fig4}, respectively.
The scatter plots
correspond to three values of the
lightest neutrino mass:
$i)$ $m_1= 3.8\times 10^{-3}$ eV, $5\times 10^{-3}$ eV and
$7\times 10^{-3}$ eV (NO spectrum);
$ii)$ $m_3=0.02$ eV, 0.06 eV and 0.1 eV (IO spectrum).
In all numerical calculations we neglect
the RG effects on neutrino masses and mixings.
This is a sufficiently
good approximation provided the light
neutrino mass spectrum is not QD \cite{RGnuref}.
In the class of $A_4$ models we consider,
the latter condition is fulfilled for the
NO spectrum since the lightest neutrino
mass $m_1$ is constrained to lie in the interval
$(3.8 \div 7) \times 10^{-3}$ eV.
In the case of IO spectrum,
the condition is approximately
satisfied for $m_3 \lesssim 0.1$ eV.
Correspondingly, for IO spectrum
we present results for
$0.02~{\rm eV}\lesssim m_3 \lesssim 0.1$ eV.

\subsection{Predictions of the AF model}

 The results for the AF model are shown in Fig. \ref{Fig3}.
In the case of NO spectrum  (left panels in Fig. \ref{Fig3}),
the normalized branching ratios
$B'(\mu\rightarrow e+\gamma)$ and
$B'(\tau\rightarrow e+\gamma)$,
defined in eq. (\ref{BRprime}),
are approximately the same,
as the analysis performed in Section 3 suggested.
The branching ratios are larger for smaller
values of the lightest neutrino mass $m_1$,
the dominant contribution being due
to the term $\propto\log(m_2/m_1)$
which is a decreasing function of  $m_1$
(Fig. \ref{Fig2}, left panel).
The same feature is exhibited by the term
$\propto \log(m_3/m_1)$. The latter is
multiplied by a coefficient of
$\mathcal{O}(\e)$.
As we have already indicated,
the term $\propto\log(m_1/m_*)$
in the AF model is suppressed,
being of $\mathcal{O}(\e^2)$,
and has a negligible effect on the results.
Due to the fact that the coefficient of the term $\propto \log (m_3/m_1)$
in $B^\prime(\tau\rightarrow \mu+\gamma)$ is of order one, the
normalized branching ratio of $\tau\rightarrow\mu +\gamma$
decay is approximately by a factor of ten
larger than those of
$\mu \rightarrow e+\gamma$ and $\tau \rightarrow e+\gamma$
decays, which
is consistent with the
analytic estimates given in
eqs. (\ref{BRNOest1}) and (\ref{BRNOest2}).

    We observe that, for the set of boundary conditions
we have chosen, eq.  (\ref{set}),
the MEGA upper limit \cite{MEGA} on
$B(\mu\rightarrow e+\gamma)$ is not satisfied for
$m_1= 3.8\times10^{-3}$ eV.
 This important experimental constraint
can be satisfied
for larger values of the lightest
neutrino mass and, in particular,
for the two other chosen values
of $m_1$, $m_1= 5\times 10^{-3}$ eV and
$m_1=7\times 10^{-3}$ eV.
However, $B(\mu\rightarrow e+\gamma)$ is always
larger than $10^{-12}$ and thus is
within the range of sensitivity of
the MEG experiment \cite{MEG},
$B(\mu\rightarrow e+\gamma) \gtrsim 10^{-13}$,
which is currently taking data.
The predicted rates of the $\tau$ LFV
radiative decays are  always below the current
experimental upper bounds
\cite{Aubert:2009tk} as well as below
the sensitivity planned to be reached at a
SuperB factory \cite{SuperB}.

 In the case of IO mass spectrum, the
predicted $B(\mu \rightarrow e +\gamma)$ is always
compatible with the existing experimental
upper limit \cite{MEGA}.
In this case the MEG experiment
will probe a relatively large region
of the parameter space of the model.
The branching ratios of $\mu\rightarrow e +\gamma$
and $\tau\rightarrow e +\gamma$ decays are, in general,
smaller by up to two orders of magnitude
than in the case of a neutrino spectrum with NO.
As we have explained earlier,
this is partly due to the fact
that the term  $\propto \log(m_2/m_1)$,
which in the case of NO mass spectrum
gives the dominant contribution,
is strongly suppressed
since $m_2$ and $m_1$ are nearly equal,
$m_2 \cong m_1$, and partly due to
the fact that the
coefficient of the term proportional
to $ \log(m_1/m_*)$ is of order $\epsilon^2$.
This conclusion is
valid for all allowed values of the
lightest neutrino mass,
$m_3 \gtrsim 0.02$ eV
(Fig. \ref{Fig2}, right panel).
In contrast to the case of a NO neutrino mass spectrum,
the branching ratios of $\mu\rightarrow e +\gamma$
and $\tau\rightarrow e +\gamma$ decays do not
show any significant dependence on the lightest neutrino mass, $m_3$.
At the same time, the $\tau\rightarrow \mu+\gamma$
decay branching ratio exhibits
a strong dependence on
the value of $m_3$.
Indeed, it varies by up to two
orders of magnitude when $m_3$ is varied from
0.02 eV to $0.1$ eV (Fig. \ref{Fig3}, right bottom panel).
The magnitude and the behaviour of
$B^\prime(\tau\rightarrow \mu+\gamma)$ as a function of
$m_3$ is determined by the term proportional to
$\log(m_3/m_1)$ in the right-hand side of eq. (\ref{BR_logs}).
It has a maximal value for
$m_3 = 0.02$ eV and decreases
as $m_3$ increases, following the decreasing
of $\log(m_3/m_1)$. As a consequence of the suppression of
the coefficient of the $\log(m_1/m_*)$ term,
the analytic estimates reported in
eqs. (\ref{BRIOest1}) and (\ref{BRIOest2}) are
valid. Thus, the $\tau\to\mu+\gamma$ decay has a branching ratio
which, at least for $m_3\approx 0.02$ eV,
is by approximately
two orders of magnitude larger than
those of the two other charged LFV radiative decays.
For $m_3=0.02$ eV we have
$B'(\tau\rightarrow \mu+\gamma)\approx 10^{-10}$.
Therefore as like in the case of NO spectrum,
the predicted $B'(\tau\rightarrow \mu+\gamma)$
for the values of mSUGRA parameters considered
is below the sensitivity range of
the currently planned experiments.

\subsection{Predictions of the AM model}

Our results for the AM model and  both types
of neutrino mass spectrum are
illustrated in Fig. \ref{Fig4}.
As was discussed above, the main difference with
the AF model is in  the prediction for
the coefficient of the term
$\log(m_1/m_*)$. In the AM model this coefficient
is of $\mathcal{O}(\e)$ for the three radiative
decays and, therefore, the term
$\propto \log(m_1/m_*)$ is not negligible.
Obviously, $\log(m_1/m_*)$ is a monotonically
increasing function of the lightest neutrino mass
(Fig. \ref{Fig2}).
Since the coefficient of this logarithm is
a number with absolute value
of order one, for both types of neutrino mass spectrum
the $\mu\rightarrow e+\gamma$,
$\tau\rightarrow e+\gamma$ and
$\tau\rightarrow \mu+\gamma$ decay
branching ratios exhibit much weaker dependence on
the lightest neutrino mass compared to the dependence
they show in the AF model.
Most importantly, as a consequence of the
contribution due to the term $\propto \log(m_1/m_*)$,
$B'(\mu \rightarrow e +\gamma)$ and
$B'(\tau\rightarrow e +\gamma)$
in the case of IO
spectrum are predicted to be of the same order
of magnitude as in the case of
NO spectrum (Fig. \ref{Fig4}).
This is in sharp contrast to the predictions of
the AF model.

 We find that the predictions
for  $B'(\tau \rightarrow \mu +\gamma)$ in the cases of
NO and IO spectrum essentially do not differ and are
similar to those obtained in the AF model.
As Fig. \ref{Fig4} shows, for both the NO and IO
mass spectrum we get $B(\mu\rightarrow e+\gamma) < 1.2\times 10^{-11}$
in roughly half of the parameter space
explored. At the same time, in practically
all the parameter space considered we find that
$B(\mu\rightarrow e+\gamma) \gtrsim 10^{-13}$.
 The $\tau$ LFV radiative decays are predicted
to proceed with rates which are below the
sensitivity range of the planned experiments.


\mathversion{bold}
\subsection{On the Possibility of Large
$B(\mu\rightarrow e+\gamma) > 10^{-13}$ and
$B(\tau\rightarrow\mu+\gamma)\approx 10^{-9}$
in the AF Model}
\mathversion{normal}


\begin{figure}[t!!]
\begin{center}
\begin{tabular}{cc}
\includegraphics[width=7.5cm,height=6.5cm]
{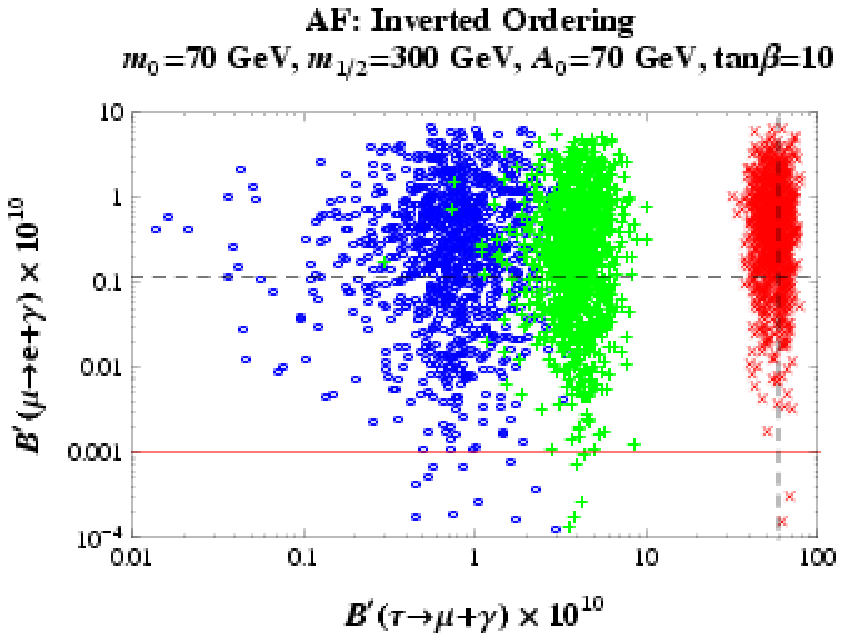}
\includegraphics[width=7.5cm,height=6.5cm]
{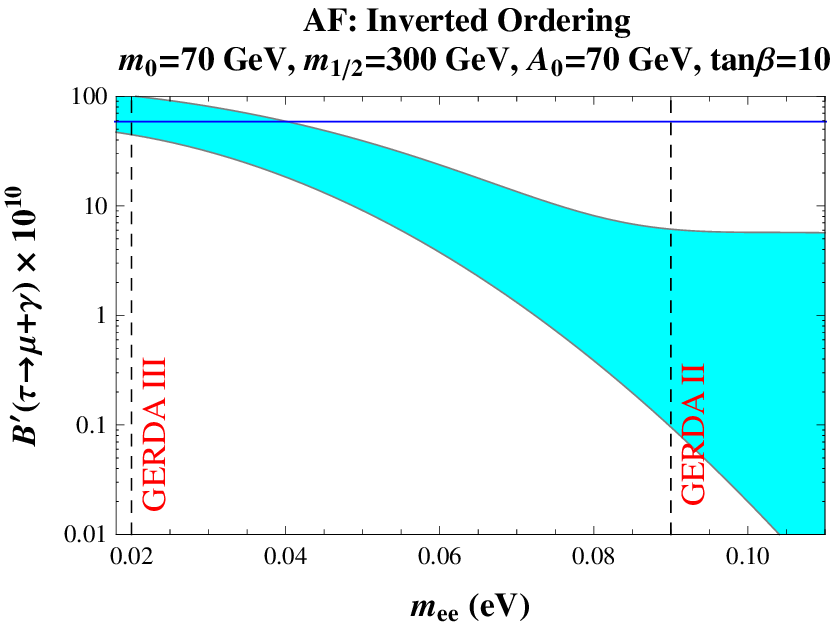}
\end{tabular}
 \caption{Left panel: correlation between
$B'(\mu\rightarrow e+\gamma)$ and $B'(\tau\rightarrow \mu+\gamma)$
in the AF model for three different values of the
lightest neutrino mass: $m_3=0.02$ eV (red $\times$),
$m_3=0.06$ eV (green $+$) and  $m_3=0.1$ eV
(blue $\circ$). The horizontal dashed line shows the current upper bound from the MEGA
experiment \cite{MEGA}, while the continuous line
corresponds to the foreseen sensitivity of the MEG experiment \cite{MEG}.
 The  vertical dashed line indicates the possible future bound on
$\tau\to\mu+\gamma$ from a SuperB factory \cite{SuperB}.
 Right panel: correlation between $B'(\tau\rightarrow \mu+\gamma)$
and the effective Majorana mass $m_{ee}$. The  horizontal continuous
line shows the prospective reach of a SuperB factory. The two dashed
vertical lines indicate the expected sensitivity of the GERDA II and
GERDA III phase \cite{GERDA}.
\label{Fig5}
}
\end{center}
\end{figure}
%

 As we have seen, for the point in the mSUGRA parameter space
considered the $\tau\rightarrow e+\gamma$ and
$\tau\rightarrow \mu+\gamma$ decay branching ratios
are predicted to be compatible with
the existing experimental upper bounds and
below the sensitivity of the future planned
experiments. However,
the decay $\tau\rightarrow \mu+\gamma$
might have a rate within the sensitivity range
of the future experiments if
the SUSY particle masses are smaller
(i.e., the effective SUSY mass scale
$m_S$, eq. (\ref{mS}) is lower)
than those resulting from eq. (\ref{set}).
This possibility can be realized
for smaller values (than those we have employed)
of the mass parameters $m_0$ and $m_{1/2}$.
Indeed, consider the following set of values:
\begin{eqnarray}
 && m_0=70\,\GeV,\,\,\,\,m_{1/2}=300\,\GeV,\,\,\,\,A_0=70\,\GeV,\,\,\,\,\tan\beta=10.\label{set2}
\end{eqnarray}
%
For the values given in eq. (\ref{set2})
squarks can be as light as $500$ GeV,
gluinos have masses
of approximately $700$ GeV and all sleptons
have masses smaller than $250$ GeV. The LSP providing the
correct amount of DM in the Universe is bino-like and
has a mass of $115$ GeV. The parameters given in eq. (\ref{set2})
correspond also to a point in the stau co-annihilation region,
very close to the region excluded by the LEP2 data
\cite{Baer:2003wx,Ellis:2003cw,Baer:2009vr}:
 the mass of the lightest Higgs boson is near $114.4$ GeV
\footnote{All masses have been calculated again by using
the program ISAJET 7.69 \cite{isajet}.}.
For the indicated values of the SUSY breaking
parameters the predicted LFV branching ratios are larger
than those corresponding to the mSUGRA point in eq. (\ref{set})
since $B_0 (m_0,m_{1/2})\tan^2\beta\approx 2.3\times 10^{-8}$.
As a result, the AM model is strongly disfavored by the
experimental limit on $B(\mu\rightarrow e+\gamma)$.
In the AF model the latter constraint cannot be satisfied,
if the neutrino mass spectrum is with NO.
In the case of IO mass spectrum, however,
the predicted $B(\mu\rightarrow e+\gamma)$
is compatible with the MEGA bound
in nearly half of the region of the
relevant parameter space
and (with the exception of
singular specific points)
is within the sensitivity
reach of the MEG experiment.
We show in Fig. \ref{Fig5}, left panel, the correlation between
the normalized branching ratios of the decays $\mu\rightarrow e+\gamma$ and
$\tau\rightarrow \mu+\gamma$ in
the AF model, assuming IO light neutrino
mass spectrum. The prospective sensitivity
of the searches for the
$\tau\rightarrow\mu+\gamma$ decay, which can be
reached at a SuperB factory,
$B(\tau\rightarrow\mu+\gamma)\approx 10^{-9}$ \cite{SuperB},
is also indicated. Assuming a scenario in which
in the MEG experiment it is found that
$B(\mu\rightarrow e+\gamma) > 10^{-13}$
and the SUSY particles with masses, as predicted above,
are observed at LHC, we see from Fig. \ref{Fig5}, left panel, that
$B(\tau\rightarrow\mu+\gamma)$ might be
detectable at a SuperB factory if the
lightest neutrino mass $m_3\approx 0.02$ eV.
 For $m_3 = 0.02$ eV, the
$\betabeta$-decay
effective Majorana mass  is predicted \cite{HMP09} to lie in the
interval $m_{ee}\approx (0.018 \div 0.054)$ eV. Values of $m_{ee}$ in the indicated interval might be probed
in some of the next generation of $\betabeta$-decay
experiments (see, e.g. \cite{GERDA, CUORE}).
 In Fig. \ref{Fig5}, right panel, we show the correlation
between the normalized branching ratio of $\tau\to \mu+\gamma$ decay and
the effective Majorana mass $m_{ee}$.
The relation between $m_{ee}$ and the lightest light neutrino mass is
discussed in \cite{AM,HMP09,BDiBFN09}:
$m_{ee} \cong \sqrt{m_3^2+|\dma|}\left|2 + e^{i\alpha_{21}}\right|/3$, where
$m_3$ and $\alpha_{21}$ are both functions of one
parameter and thus their values are correlated.
We indicate the prospective
sensitivity of the GERDA II and GERDA III phase,
$m_{ee}=0.09$ eV and $m_{ee}=0.02$ eV,
respectively \cite{GERDA}. As one can see,
with a positive
signal of $B(\tau\to \mu+\gamma) \approx 10^{-9}$ at a SuperB factory
values of $m_{ee}$ up to $m_{ee} \approx 0.04$ eV can be probed.
The sum of neutrino masses reads \cite{HMP09}
$\sum m_i \approx 0.125$ eV for $m_3 \approx 0.02$ eV. This value is
smaller than the current cosmological bounds
(see, e.g. \cite{Fogli}), but is within the
sensitivity expected to be reached
by combining data on weak lensing
of galaxies by large scale
structure with data from WMAP and
PLANCK experiments (see, e.g. \cite{Hann06}).

\mathversion{bold}
\subsection{Specific Features of the Predictions for $B(\mu\to e+\gamma)$}
\mathversion{normal}

\begin{figure}[t!!]
\begin{center}
\begin{tabular}{cc}
\includegraphics[width=7.5cm,height=6.5cm]{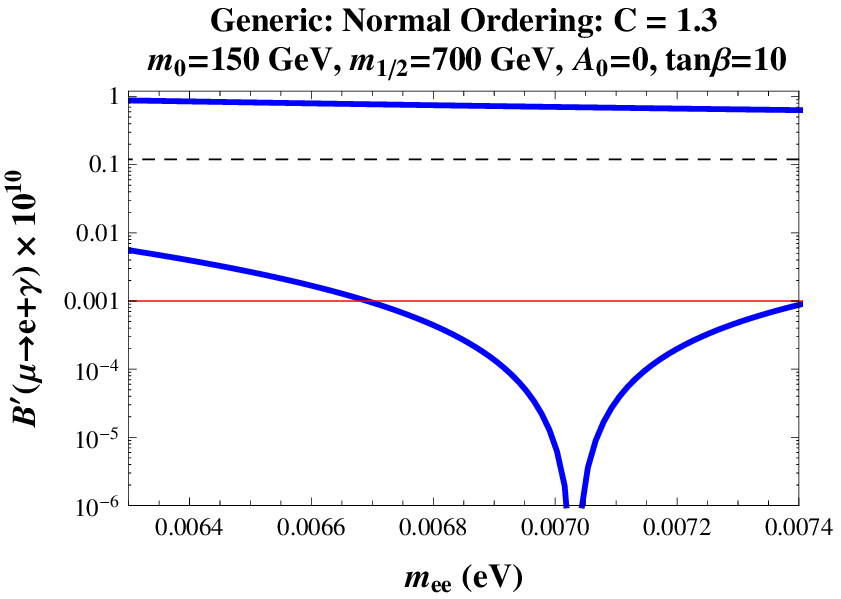} &
\includegraphics[width=7.5cm,height=6.5cm]{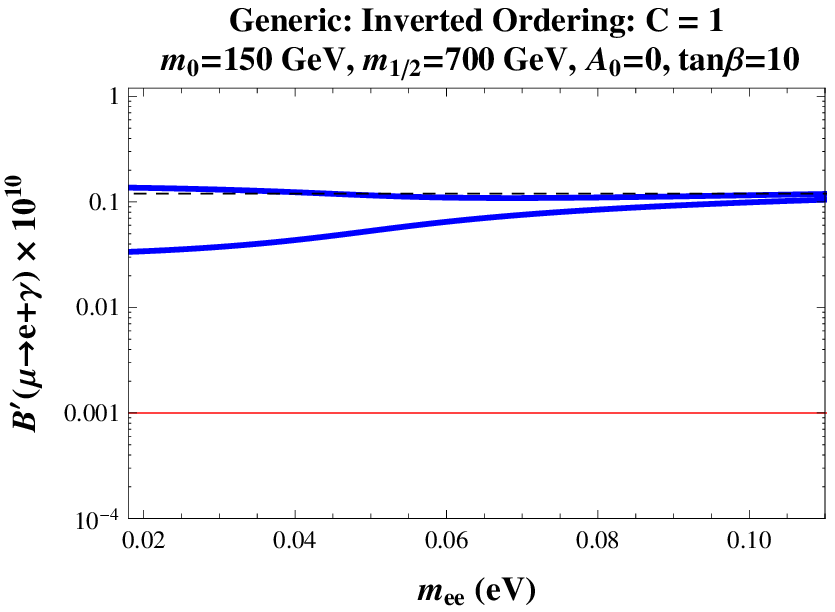}
\end{tabular}
\caption{$B'(\mu\to e+\gamma)$ vs $m_{ee}$ for NO (left panel) and
IO (right panel) light neutrino mass spectrum calculated
for an $A_4$ model with generic NLO corrections, see eq.(\ref{deltamD}).
 Lower and upper limits on
$B'(\mu\to e+\gamma)$ are shown, which can be
found by using eq. (\ref{BRC}) for all possible
combinations of $\sigma_{1,2,3}$. The horizontal dashed line
corresponds to the MEGA bound \cite{MEGA}, $B'(\mu\rightarrow e
+\gamma)\leq 1.2 \times 10^{-11}$. The horizontal continuous line
corresponds to $B'(\mu\rightarrow e +\gamma) = 10^{-13}$, which is
the prospective sensitivity of the MEG experiment \cite{MEG}. The
results shown correspond to the best fit values \cite{nudata}
$|\dma|=2.40\times10^{-3}$ eV$^2$ and $r=\dmsol/|\dma|=0.032$.
\label{Fig6}}
\end{center}
\end{figure}
%

Apart from discussing generic results for the AF and the AM model
it is also interesting to have a closer look at particular points
in the parameter space of the $A_4$ models (with generic NLO
corrections). In order to do so we use the analytic
formula given in Section 3.2., eq.(\ref{BR_logs}),
for the branching ratio of the decay $\mu\to e+\gamma$
together with the results given in Table 1 and assume that the
coefficients of the $\mathcal{O}(\epsilon)$ terms are real and
have the same absolute value $C>0$:
\begin{equation}
B'(\mu\rightarrow e+\gamma)\;\propto\; \left|\frac{1}{3} y_\nu^2
\log\left(\frac{m_2}{m_1}\right) + C \, \epsilon \, \left(
\sigma_1 \log\left(\frac{m_1}{m_*}\right) + \sigma_2
\log\left(\frac{m_2}{m_1}\right) + \sigma_3
\log\left(\frac{m_3}{m_1}\right) \right) \right|^2  \; .
\label{BRC}
\end{equation}
%
We do not fix the relative sign of these terms and allow
for all eight combinations $\sigma_{1,2,3}= \pm 1$. We choose for
the mSUGRA parameters the values given in eq. (\ref{set}), set $\epsilon=0.04$, set
$y_\nu=1$ and take again best fit values for $r$ and $|\dma|$. In
Fig. \ref{Fig6}, left panel, we plot the result for $B(\mu\to
e+\gamma)$ for $C=1.3$ in the case of a NO light neutrino mass
spectrum with respect to the effective Majorana mass $m_{ee}$,
$m_{ee} \cong |2\,m_1 + \sqrt{m_1^2+\dmsol}|/3$.
We show only the two curves which correspond to the upper and
lower bound that can be reached for
the eight different combinations of $\sigma_{1,2,3}$.
As one can
see, there exists the possibility of cancellations between the
terms contributing to the branching ratio of
the $\mu\to e+\gamma$ decay, so
that the value of the latter can be strongly suppressed
\footnote{Note that the value of $B(\mu\to e+\gamma)$ will still
be non-zero in general, because we expect corrections to the
coefficients of the different logarithms of order
$\epsilon^2$.}.
The value of $m_{ee}$ at which the suppression takes
place depends on the value of $C$.
In Fig. \ref{Fig6}, right panel, we
show the corresponding plot for IO
neutrino mass spectrum. We choose $C= 1$.
In contrast to the case of NO spectrum,
no strong suppression of $B(\mu\to e+\gamma)$
is possible, because the term $\propto \log(m_1/m_*)$
 always dominates (see Fig.
\ref{Fig2}, right panel).
 This result holds for all values of
the constant $C$ from the interval $0.1 \lesssim C \lesssim 6$.
 Allowing for arbitrary relative phases between
the different contributions in the
right-hand side of eq. (\ref{BRC}), we find that for a NO light
neutrino mass spectrum the curve for $\sigma_{1,2,3}=+1$
corresponds to an upper bound on $B(\mu\to e +\gamma)$, whereas
the curve for $\sigma_{1,2,3}=-1$ is an absolute lower bound with the
exception of few points in the parameter space.
For the IO spectrum,
the bounds obtained for real coefficients are also
upper and lower bounds in the case of arbitrary relative phases between
the different terms in eq. (\ref{BRC}).

  As mentioned earlier, the preceding analysis holds for
an $A_4$ model with generic NLO corrections, as
is the case of the AM model.
In order to perform a similar analysis for the AF model
\footnote{ We remind the reader that
in the AF model the coefficient of the logarithm
$\log (m_1/m_*)$ is of order $\epsilon^2$ rather than
$\epsilon$.},
we replace $\sigma_1 \log (m_1/m_*)$ with
$\epsilon \, \sigma_1 \log (m_1/m_*)$
in eq. (\ref{BRC}). We find that
deep cancellations between the different contributions
in $B(\mu\to e +\gamma)$ can occur
in both the cases of NO and IO neutrino mass spectrum.
For the IO spectrum, the cancellations leading to a strong suppression of
branching ratio $B(\mu\to e +\gamma)$
take place for $m_{ee}$ around 0.09 eV
for almost all values of $C$ in the range
considered, $0.1 \lesssim C \lesssim 6$.

\mathversion{bold}
\section{The $\mu-e$ Conversion and $\ell_i\to3\ell_j$ Decay Rates}
\mathversion{normal}

We briefly discuss in this section the experimental constraints
that can be imposed on the $A_4$ models from data on
$\mu- e$ conversion and the decays $\ell_i\to 3\ell_j$.
In the mSUGRA scenario, these LFV processes are
dominated by the contribution coming from
the $\gamma-$penguin diagrams.
As a consequence, for $\mu-e$ conversion,
the following relation holds with a good approximation \cite{Hisano96}:
\begin{equation}
 CR(\mu\, {\rm N}\to e\, {\rm N})\equiv\frac{\Gamma(\mu {\rm N}\to e {\rm N})}{\Gamma_{\rm capt}}=\frac{\alpha_{em}^4 G_F^2 m_\mu^5 Z}{12 \pi^3 \Gamma_{\rm capt}}Z_{eff}^4|F(q^2)|^2 B(\mu\to e+\gamma)\; .\label{mu2e}
\end{equation}
%
 In eq. (\ref{mu2e}) $Z$ is the proton number in the nucleus N,
$F(q^2)$ is the nuclear form factor at momentum transfer $q$,
$Z_{eff}$ is an effective atomic charge and
$\Gamma_{\rm capt}$ is the experimentally known
total muon capture rate.
For $_{22}^{48}{\rm Ti}$ we have $Z_{eff}=17.6$,
$F(q^2=-m_{\mu}^2)\approx 0.54$ and
$\Gamma_{\rm capt}=2.590\times 10^6\,{\rm sec^{-1}}$ \cite{PRIME}.
In the  case of $ _{13}^{27}{\rm Al}$
one finds $Z_{eff}=11.48$, $F(q^2=-m_{\mu}^2)\approx 0.64$
and $\Gamma_{\rm capt}=7.054\times 10^5\,{\rm sec^{-1}}$ \cite{Mu2eColl}.
According to eq. (\ref{mu2e}), the $\mu - e$ conversion ratios
in $_{22}^{48}{\rm Ti}$ and $ _{13}^{27}{\rm Al}$ are given by:
\begin{equation}
 CR(\mu\, _{22}^{48}{\rm Ti} \to e\, _{22}^{48}{\rm Ti})\approx 0.005 B(\mu\to e+\gamma)
\; , \;\; CR(\mu\, _{13}^{27}{\rm Al} \to e\, _{13}^{27}{\rm Al})\approx 0.0027 B(\mu\to e+\gamma) \; .
\end{equation}
%
\indent Future experimental searches
for $\mu-e$ conversion can reach
the sensitivity:
$CR(\mu\, _{22}^{48}{\rm Ti} \to e\, _{22}^{48}{\rm Ti})
\approx 10^{-18}$ \cite{PRIME},
and $CR(\mu\, _{13}^{27}{\rm Al} \to e\, _{13}^{27}{\rm Al})\approx 10^{-16}$
\cite{Mu2eColl}. The upper bound $B(\mu\to e+\gamma)<10^{-13}$
which can be obtained in  the MEG experiment
would correspond to the following upper bounds on the
$\mu - e$ conversion ratios under discussion:
$CR(\mu\, _{22}^{48}{\rm Ti} \to e\, _{22}^{48}{\rm Ti})
< 5 \times 10^{-16}$ and
$CR(\mu\, _{13}^{27}{\rm Al} \to e\, _{13}^{27}{\rm Al})
< 2.7 \times 10^{-16}$. The latter could be probed
by future experiments on $\mu-e$ conversion,
which have higher prospective sensitivity.
We see that for the mSUGRA points considered
in Section 4,
both the AF and AM models
can be further constrained
by the experiments on $\mu - e$ conversion
if the $\mu\to e+\gamma$ decay will not be
observed in the MEG experiment.

  In what concerns the decay of a charged lepton
into three charged leptons, the branching ratio
is approximately given by \cite{Hisano96}:
\begin{equation}
 B(\ell_i\to3\ell_j)\approx
\frac{\alpha}{3\pi}\left(\log\left(\frac{m_{\ell_i}^2}{m_{\ell_j}^2}\right)-\frac{11}{4}\right)B(\ell_i\to\ell_j+\gamma)
\; .
\end{equation}
%
The searches for $\tau \to \mu + \gamma$,
$\tau \to 3\mu$ and $\tau \to 3e$
decays at SuperB factories \cite{SuperB}
will be sensitive to
$B(\tau\to \mu + \gamma),B(\tau\to3\mu),B(\tau\to 3 e)\geq 10^{-9}$.
Therefore, if in the experiments at SuperB factories
it is found that $B(\tau\to \mu + \gamma) <  10^{-9}$,
obtaining the upper limits
$B(\tau\to3\mu),B(\tau\to 3 e) < 10^{-9}$
would not constrain further
the $A_4$ models considered here.
However, the observation of the $\tau\to3\mu$ decay
with a branching ratio $B(\tau\to3\mu) \geq 10^{-9}$,
combined with the upper limit
$B(\tau\to \mu + \gamma) <  10^{-9}$, or the
observation of the
$\tau\to \mu + \gamma$ decay
with a branching ratio $B(\tau\to \mu + \gamma) \geq 10^{-9}$,
would rule out the $A_4$ models under discussion.

 The current limit on the $\mu\to 3 e$ decay branching ratio is
$B(\mu \to 3e) < 10^{-12}$ \cite{SINDRUM}. There are no plans at present to perform
a new experimental search for the $\mu\to 3 e$ decay with higher
precision.


\section{Conclusions}


  In this paper we have studied charged lepton
flavour violating (LFV) radiative decays,
$\mu\rightarrow e+\gamma$,
$\tau\rightarrow \mu+\gamma$ and
$\tau\rightarrow e +\gamma$
in a class of supersymmetric $A_4$ models
with three heavy RH Majorana neutrinos,
in which the lepton (neutrino) mixing is
predicted to leading order (LO) to be tri-bimaximal (TB).
The light neutrino masses are generated
via the type I see-saw mechanism.
We work within  the framework of the minimal
supergravity (mSUGRA) scenario,
which provides flavour universal boundary conditions
at the scale of grand unification
$M_X \approx 2 \times 10^{16}$ GeV.
In this scenario the flavour universal
slepton masses, trilinear couplings and the gaugino masses
are characterised by the parameters $m_{0}$, $A_0$ and $m_{1/2}$
respectively. The other free parameters are $\tan\beta$
and $\mbox{sign}(\mu)$.
Flavour off-diagonal
elements in the slepton
mass matrices are
generated through RG effects
(from the scale $M_X$ to the
scale of heavy RH Majorana
neutrino masses).
The former induce the LFV decays $\mu\to e+\gamma$,
$\tau\to\mu+\gamma$ and $\tau\to e+\gamma$.
In the present article
the branching ratios $B (\ell_i \to \ell_j + \gamma)$
are calculated using the
analytic approximations
developed in \cite{Hisano96,BR1,BR2}.
In this approach $B (\ell_i \to \ell_j + \gamma)$
depend only on the
matrix of neutrino Yukawa couplings $\hat{Y}_\nu$
defined in the flavour basis, on the
three heavy RH Majorana
neutrino masses, and on an ``average''
SUSY mass scale, $m_S$ (see eq. (\ref{mS})).
In the class of $A_4$ models considered,
each of the three light Majorana
neutrino masses $m_i$ is
directly related at LO to the
corresponding heavy Majorana neutrino
mass $M_i$, eq. (\ref{numass}).
Both types of light neutrino mass spectrum (with
normal ordering (NO) and inverted ordering (IO)),
are possible. The lightest
neutrino masses, corresponding to the two types
of spectrum, are constrained to lie in the intervals:
$m_1 \approx (3.8 \div 7.0)\times 10^{-3}$ eV (NO) and
$m_3 \gtrsim 0.02$ eV (IO).
The  RG effects on
the neutrino masses and
mixing angles were not taken into
account.
As is well known, these effects are relatively
small if the lightest neutrino mass satisfies
${\rm min}(m_j) < 0.10$ eV.
For this reason we limited
our analysis in the case of IO spectrum
to $m_3 \lesssim 0.10$ eV.

 The analytic estimates of the branching ratios
$B (\ell_i \to \ell_j + \gamma)$
we have given were made for the case of
generic next-to-leading order
(NLO) corrections to the neutrino
Yukawa matrix. The numerical results
we presented, however, are obtained for two
explicit realizations of the $A_4$ models, those
by Altarelli and Feruglio (AF) \cite{AF2} and
by Altarelli and Meloni (AM) \cite{AM}, respectively.
In these models the $A_4$ symmetry is
broken spontaneously at high energies and
the symmetry breaking parameter $\epsilon$
has a value in the interval
$0.007 \lesssim \epsilon \lesssim 0.05$.
In the numerical calculations
we set to $\epsilon=0.04$.
In this case the allowed range of
$\tan\beta$
is $2 \lesssim \tan\beta \lesssim 12$.
We used the value of
$\tan\beta = 10$ in the numerical calculations.

  The predictions for the $\ell_i \to \ell_j + \gamma$
decay branching ratios, $B(\ell_i \to \ell_j + \gamma)$,
in the AF and AM models
are derived for one specific point in
the mSUGRA parameter space
lying in the stau co-annihilation region,
which is compatible with direct bounds on
sparticle masses and the requirement of explaining
the amount of dark matter (DM) in the Universe:
$m_0=150$ GeV, $m_{1/2}=700$ GeV and $A_0=0$ GeV.
These values correspond to
a bino-like LSP with a mass of approximately
$280$ GeV,
which makes up the DM of the Universe,
gluino masses of $1.6$ TeV and
squark masses between $1.1$ TeV and $1.5$ TeV.
Having the indicated masses,
the  SUSY particles can be observed at LHC.
Results for other points in the  mSUGRA
parameter space can be easily
obtained by modifying the rescaling function
present in the expressions for $B(\ell_i \to \ell_j + \gamma)$
(see eqs. (\ref{BR}) and (\ref{B0})).

  We have found that in the case of NO light neutrino mass
spectrum, both the AF and AM models predict
$B(\mu \rightarrow e +\gamma) > 10^{-13}$
in practically all the parameter space considered
(Figs. \ref{Fig3} and \ref{Fig4}). The same conclusion
is valid for the IO mass spectrum in the case of
the AM model, whereas for the AF model
this result holds roughly in half
of the parameter space of the model.
Values of $B(\mu \rightarrow e +\gamma) \gtrsim 10^{-13}$
can be probed in the MEG experiment which is taking
data at present.

The predictions of the AF model
for all the three branching ratios
$B(\ell_i \to \ell_j + \gamma)$ in the case of
NO spectrum show a noticeable
dependence on the value of the
lightest neutrino mass.
 The dependence of $B(\tau\to \mu +\gamma)$
on ${\rm min}(m_j)$ is particularly strong
in the case of IO spectrum.
 In contrast, $B(\mu\to e +\gamma)$
and $B(\tau\to e +\gamma)$ in this case
vary relatively little with ${\rm min}(m_j)$.
  The predictions for $B(\ell_i \to \ell_j + \gamma)$
in the AM model do not exhibit significant
dependence on ${\rm min}(m_j)$.
The branching ratios
$B(\tau\to e +\gamma)$ and $B(\tau\to \mu +\gamma)$
are always predicted to be below the sensitivity
of the present and future planned experiments.
We have shown, however, that if the
SUSY particles are lighter,
 e.g. the LSP has a mass of $115$ GeV,
squarks are as light as $500$ GeV
and gluinos have masses $\approx 700$ GeV,
one can have $B(\mu \rightarrow e +\gamma) \gtrsim 10^{-13}$
and $B(\tau\to \mu +\gamma)
\approx 10^{-9}$ in the AF model with
IO spectrum (Fig. \ref{Fig5}, left panel). A value of
$B(\tau\to \mu +\gamma) \approx 10^{-9}$
requires the lightest neutrino mass to be
$m_3 \approx 0.02$ eV.
Sensitivity to  such a value of
$B(\tau\to \mu +\gamma)$
can be achieved, in principle,
at a SuperB factory \cite{SuperB}.

   We found that the dependence of the
branching ratios $B(\ell_i\rightarrow \ell_j+\gamma)$
on the $A_4$ flavour symmetry breaking parameter
$\epsilon$ in the case of NO
spectrum  differs from that in the case
of IO spectrum. For the NO spectrum,
all the three branching ratios
are predicted to scale as $\epsilon^0$.
The  normalized branching ratios
$B'(\mu\to e +\gamma) \approx B(\mu\to e +\gamma)$ and
$B'(\tau\to e +\gamma) =
B(\tau\to e +\gamma)/B(\tau\to e +\nu_\tau + \bar{\nu}_e)$,
are found to be equal at LO,
while $B'(\tau\to \mu +\gamma)\equiv
B(\tau\to \mu +\gamma)/B(\tau\to \mu +\nu_\tau + \bar{\nu}_{\mu}) $
is typically larger by one order of magnitude.
 In the case of IO spectrum we have at LO
$B(\mu\to e +\gamma) \propto \epsilon^2$,
$B'(\tau\to e +\gamma) \propto \epsilon^2$.
However, for $\epsilon\approx 0.04$,
this $\epsilon$ suppression is compensated
by relatively large values of the logarithm
$\log (m_1/m_*)$. Therefore in a model
with generic NLO corrections to the
matrix of neutrino
Yukawa couplings, $B(\mu\to e +\gamma)$ and
$B'(\tau\to e +\gamma)$ can have the same
magnitude for the two types - NO and IO,
of neutrino mass spectrum.
 The
specific structure of the NLO
contributions to the matrix of neutrino
Yukawa couplings in the AF model
leads to an additional suppression of
the branching ratios $B(\mu \rightarrow e +\gamma)$
and $B'(\tau \rightarrow e +\gamma)$.
As a consequence,
$B(\mu \rightarrow e +\gamma)$
and $B'(\tau \rightarrow e +\gamma)$
in the case of IO spectrum
are predicted to be smaller
in the AF model than  in the AM model
by up to two orders of magnitude.
Since $B'(\tau\to\mu+\gamma)$
 always receives a contribution
of order $\epsilon^0$, it has the
same magnitude for the NO and
IO spectrum, and in the case of IO spectrum
and for  $m_3 \approx 0.02$ eV
it is not affected by the indicated
additional suppression.
 For the AF model thus
$B'(\tau\to\mu+\gamma)$ can be by
up to three orders of magnitude
larger than $B(\mu\to e+\gamma)$ and
$B'(\tau\to e+\gamma)$.

  The type of dependence of the branching ratios
on the $A_4$ symmetry breaking parameter
$\epsilon$ is in contrast to that
found in the framework of an effective field
theory approach in \cite{EFTA4}, in which the setup of the
AF model was used. This approach
 suggests that $B(\ell_i\rightarrow \ell_j+\gamma)$ scale
generically as $\epsilon^2$, and as  $\epsilon^4$ in a specific
supersymmetric version of the theory.
Moreover, the results found in the effective field theory approach
do not show any dependence on the type of the light neutrino
mass spectrum and on the value of the lightest neutrino mass.

 In \cite{HMP09} we studied versions of the AF and AM models,
in which the neutrino Yukawa couplings are suppressed,
being proportional to $\epsilon$. As a consequence,
the scale of the RH neutrino masses is lowered.
Since the branching ratios of the charged LFV
radiative decays studied in the present article
are proportional to the Yukawa couplings to the
forth power, suppressing the latter
also efficiently suppresses the
$\ell_i\rightarrow \ell_j+\gamma$
decay branching ratios.
Thus, for the mSUGRA points
chosen in our numerical
study and for most of the relevant
parameter space of the version of $A_4$ models
considered in \cite{HMP09} we expect
$B(\mu\to e+\gamma)$ to be below the
sensitivity of the on-going
MEG experiment.

We have shown that for specific values of the parameters in an $A_4$
model with generic NLO corrections, ``accidental'' cancellations between
the different contributions in the $\mu\to e + \gamma$ decay rate
and a strong suppression of the branching ratio $B(\mu\to e+\gamma)$
are possible in the case of NO neutrino mass spectrum.
The same result holds in the AF model for both types - NO and IO,
of spectrum.  No similar suppression is found
to occur for the IO spectrum in the $A_4$
model with generic NLO corrections.

 Our estimates of the predicted rate of $\mu -e$
conversion in the $A_4$ models considered show that
future experiments can further constrain
these models
if the $\mu\to e+\gamma$ decay will not be
observed in the MEG experiment.
The observation at the SuperB factories
of the $\tau\to3\mu$ decay
with a branching ratio $B(\tau\to3\mu) \geq 10^{-9}$,
combined with the upper limit
$B(\tau\to \mu + \gamma) <  10^{-9}$, or the
observation of the
$\tau\to \mu + \gamma$ decay
with branching ratio $B(\tau\to \mu + \gamma) \geq 10^{-9}$,
would rule out the $A_4$ models under discussion.
If $B(\tau\to \mu + \gamma)$ is found to satisfy
$B(\tau\to \mu + \gamma) <  10^{-9}$,
the prospective sensitivity of SuperB factories to
the decay modes $\tau\to 3 \mu$ and $\tau\to 3 e$
would not allow to obtain
additional constraints on
the parameter space of the $A_4$ models from
non-observation of the $\tau\to 3 \mu$ and $\tau\to 3 e$
decays.

    The results of the MEG experiment which is taking data
at present, and of the upcoming experiments at LHC
can provide significant tests of, and
can severely constrain, the class of
$A_4$ models  predicting tri-bimaximal
neutrino mixing, in general, and
the Altarelli-Feruglio and
Altarelli-Meloni models, in particular.

\section*{Acknowledgments}
We would like to thank L. Calibbi and Y. Takanishi for discussions.
This work was supported in part by the Italian INFN
under the program ``Fisica Astroparticellare'' and
by the European Network ``UniverseNet'' (MRTN-CT-2006-035863).
The work of S.T.P. was supported in part also
by World Premier International Research Center
Initiative (WPI Initiative), MEXT, Japan.

\end{document}